\documentclass[prd,nofootinbib, superscriptaddress,preprint]{revtex4}
\usepackage[T1]{fontenc}
\usepackage{amsmath,amssymb}
\usepackage{xcolor}
\usepackage{epsfig}
\usepackage{dcolumn}
\usepackage{subfloat}
\usepackage{graphicx}
\usepackage{csquotes}
\usepackage{slashed}
\usepackage[colorlinks,citecolor=blue]{hyperref}
\usepackage{pdfpages}
\usepackage[lofdepth,lotdepth]{subfig}
\usepackage{multirow}
\begin{document}
\title{\boldmath Linking resonant leptogenesis  with dynamics of the inverse seesaw theory with $ A_{4} $ flavor symmetry}

	\author{Maibam Ricky Devi}
	\email{deviricky@gmail.com}
	
	\author{Kalpana Bora}
	\email{kalpana@gauhati.ac.in}
	\affiliation{Department of Physics, Gauhati University, Guwahati-781014, Assam, India}

\begin{abstract}
In this paper, we analyse resonant leptogenesis in a low scale inverse seesaw model with $A_4$ flavor symmetry, in a model we explored earlier  to explain light neutrino masses and mixings, and also charged lepton flavor violating decay $\mu\rightarrow e\gamma$. Six $ A_{4} $ scalar singlets and one $ A_{4} $ fermion triplet are included, which are charged under the group  $A_{4}\times U(1)_{X} \times Z_{5} \times Z_{4} $, with at least two degenerate RH (Right Handed) neutrinos. The light neutrino masses and leptogenesis both share a same origin with the heavy right handed neutrinos. Thus, we expound the possibility of generating resonant leptogenesis in this  model at energies  as low as 1 TeV. We then analyse our findings to envision if our model inclines  more towards weak or strong washout.
\end{abstract}
	\maketitle
\section{Introduction}
\label{sec:RL1}
Explaining origin of baryon asymmetry of the universe (BAU) is one of the most important unsolved problems in physics. As per one school of thought, it is believed that the BAU has its origin from the out-of-equilibrium decay of heavy Majorana neutrinos during the early Universe. This mechanism, known as leptogenesis, is usually implemented through the so-called hierarchical scenario, which requires a hierarchical structure of the light neutrino masses and CP-violating phases. However, it is also possible to generate the BAU in a different framework, known as resonant leptogenesis (RL), which relaxes the requirement of the heavy Majorana neutrinos being hierarchical and allows a degeneracy between them. In the RL scenario, the CP-violating asymmetries produced by the heavy Majorana neutrino decays are significantly larger than in the hierarchical scenario. This is usually due to the fact that, instead of the usual one-body decay, the heavy Majorana neutrinos can decay via a two-body channel, which is strongly enhanced due to a resonant effect. Thus, the BAU can be generated even for very small values of  the CP-violating phases.  \\
\\ 
The baryon asymmetry of the universe can be  generated by  decay of the lightest right-handed neutrino which easily follows the three Sakharov's conditions \cite{Sakharov:1967dj}. To generate the lepton asymmetry, the Lagrangian needs to violate C and CP symmetries as well as the lepton number. These right-handed neutrinos can later undergo fast reactions called  \enquote{Sphaleron Process} \cite{Kuzmin:1985mm}, that converts  lepton-antilepton asymmetry to the observed matter-antimatter asymmetry. If T is the temperature of the early universe and $ M_{1} $ is the mass of the lightest right-handed neutrino then any pre-existing lepton asymmetry at $ T \geq M_{1} $, can be erased by the high energy Majorana neutrinos. However when the temperature of the universe falls below $ M_{1} $, the right-handed neutrinos can no longer follow the inverse reaction to create thermal equilibrium, leading to a surplus of heavy Majorana neutrinos. The event of leptogenesis \cite{Fukugita:1986hr} occurs at the point in the timeline of the Universe when $ T\leq M_{1} $. These right-handed Majorana neutrinos are considered to be of out-of-equilibrium when the decay width of the lightest right handed neutrino, $ \Gamma (T) $ is found to be less than the Hubble parameter $ H(T) $ that determines the expansion of the universe. When the two parameters reach an equilibrium condition, i.e., $ \Gamma (T) \cong H(T) \vert_{T=M_{1}} $, then this borderline condition is almost equivalent to the condition that the effective neutrino mass $ \widetilde{m_{i}} $ is equal to the equilibrium neutrino mass $ m_{*} $. We shall discuss this in details in the next section.\\
\\
High scale seesaw models to explain light neutrino mass are not testable, and hence low scale seesaw models are considered to be very interesting. Moreover, there is no theory so far that can authenticate the origin of Baryon Asymmetry of the Universe. With this motivation, in this work we study resonant leptogenesis \cite{Pilaftsis:1997dr, Pilaftsis:1997jf, Pilaftsis:2003gt} in the inverse seesaw model developed and presented in Ref. \cite{Devi:2021aaz,Bora:2022jdp,Devi:2022scm}, as a possible explanation of the baryon asymmetry of the universe. Low-scale seesaw models can be tested in the current and future experiments such as BASE experiment \cite{BASE:2015ity}. Some of the previous works on resonant leptogenesis at low scale can be found in \cite{Dolan:2018yqy,Dev:2015cxa,Eijima:2017anv,Drewes:2016lqo,Drewes:2016jae,Lopez-Pavon:2015cga,Boubekeur:2004ez, Chakraborty:2021azg}. In this low-scale inverse seesaw model the CP asymmetry is resonantly enhanced due to presence of two RH Majorana neutrinos of nearly degenerate masses. Testable resonant leptogenesis at GeV-TeV scale in the context of flavor symmetry and GUT has been discussed in \cite{Fong:2021tqj}, whereas some studies of gravitino constraints, which are important in the case of non-SUSY models to specify assumptions on the physical conditions in the very early Universe can be found \cite{Khlopov:1984pf}, \cite{balestra1984p}, \cite{Khlopov:1993ye},and studies of physical and cosmological effects of broken symmetry of
families, unifying PQ symmetry and mechanism of neutrino Majorana mass generation \cite{Sakharov:1994pr}, \cite{Berezhiani:1989fp}, \cite{Berezhiani:1990jj}. We consider a hierarchical structure of the light neutrino masses, with the lightest one being much smaller than the other two, and we assume that the heavy Majorana neutrinos have a quasi-degenerate spectrum. We calculate the CP-violating asymmetry produced by the heavy Majorana neutrino decays as a function of the oscillation parameters and determine the values of the oscillation parameters for which the resonance condition is satisfied. We then calculate the BAU for those values of the oscillation parameters and study the dependence of the BAU on the oscillation parameters. We compare this our results with the recent data released by the PLANCK Experiment \cite{Planck:2018vyg}. One of the interesting features of this scenario is that the resonant leptogenesis connects the neutrino physics with cosmology. The Baryon Asymmetry of the Universe can be either enhanced or suppressed for a particular set of neutrino oscillation parameters whose relevant BAU values seems to be in agreement with the observed values. Thus the neutrino oscillation parameters can significantly effect the observed Baryon Asymmetry of the Universe.\\
The paper has been organised as follows. In Section 2, we give a brief review about resonant leptogenesis. Section 3 contains details about our inverse seesaw model \cite{Devi:2021aaz,Bora:2022jdp,Devi:2022scm}
, and also about degenerate heavy RH Majorana neutrinos. Numerical analysis about the work has been presented in Section 4, and Section 5 contains results and discussion. We summarise and conclude the work in Section 6.
\section{Resonant Leptogenesis}
\label{sec:RL2}
In the scenario of resonant leptogenesis which is typically like the low-energy thermal leptogenesis, the Davidson-Ibarra bound can be circumvented when at least two of the RH neutrinos (RHN) have degenerate masses. Lepton number asymmetry can be sufficiently generated even for RH neutrino masses of $\sim$ the TeV scale provided that the self-energy effects from the RHNs get dominated in the lepton CP asymmetry. In this special case, the mass splitting between the nearly degenerate RHNs is equivalent to the decay width of the Majorana neutrinos. The Sphaleron process produces the baryon asymmetry through CP asymmetry  as they preserve the B-L quantum number.In the inverse seesaw scenario,  the resonant enhancement of the lepton asymmetry are achieved by the degenerate pseudo-Dirac neutrinos in S-N sectors, as the matrix $ \mu_{s} $ controls the small mass splitting between the two nearly degenerate RHNs. The CP asymmetry induced by the self-energy contribution in the decay of the lightest RHNs into lepton of flavor $ \alpha $ is given by  \cite{ Pilaftsis:1997jf, Pilaftsis:2003gt, Pilaftsis:1998pd, Aoki:2015owa}
\begin{eqnarray}
\epsilon_{i,j} &=&
   \frac{\sum_\alpha\left[\Gamma(N_{i,j}\rightarrow L_\alpha+H^\ast)
                        -\Gamma(N_{i,j}\rightarrow L_\alpha^c+H)\right]}
        {\sum_\alpha\left[\Gamma(N_{i,j}\rightarrow L_\alpha+H^\ast)
                        +\Gamma(N_{i,j}\rightarrow L_\alpha^c+H)\right]} 
  \nonumber \\
  &\simeq& 
   \frac{{\rm  Im}(Y_\nu^{N \dag}Y_\nu^{S})^{2}_{ij}}{(Y_\nu^{S \dag}Y_\nu^{S})_{ii}(Y_\nu^{N \dag}Y_\nu^{N})_{jj}}
   \frac{(m^{2}_{N_{i}}-m^{2}_{N_{j}})m_{N_{i}}\Gamma_{j}}{(m^{2}_{N_{i}}-m^{2}_{N_{j}})^{2}+m^{2}_{N_{i}}\Gamma^{2}_{j}}
   \label{epsilon}
\end{eqnarray}
 The decay width of the  neutrinos $N_{i,j}$ is given by $\Gamma_{i,j}=A_{i,j} m_{N_{i,j}}/(8\pi)$ where
 \begin{eqnarray}
  A_i = (Y_\nu^{S\dag}Y_\nu^S)_{ii},
  ~~~A_j=(Y_\nu^{N\dag}Y_\nu^N)_{jj}. \label{r}
 \end{eqnarray}.\\
The baryon asymmetry $ \eta_{B}$ can thus be determined from the lepton asymmetry $ \epsilon_{i,j}$ as \cite{Aoki:2015owa}
\begin{equation}
\eta_B = -\frac{28}{79}\frac{0.3 \epsilon_{i,j}}{g_{*} \kappa_{i,j} (ln \kappa_{i,j})^{0.6}} 
\end{equation}
where $g_{*}=106.75$ is the number of relativistic degrees of freedom and  the Hubble constant $H(T)=1.66 \sqrt{g_\ast}T^2/m_{\rm Pl}$, and $ m_{\rm Pl} $ is the Planck mass. The decay parameter  $\kappa $ can be used so as to estimate the expected baryon asymmetry as follows \cite{Pilaftsis:2003gt} \\
  \begin{equation}
  \kappa_{i,j} = \dfrac{\Gamma_{i,j}}{H(T)}\vert_{T=m_{N_{i,j}}} = \dfrac{\widetilde{m}}{m_{*}}
  \end{equation}
  where $ \widetilde{m} = \Sigma \widetilde{m_{l}}= \dfrac{(Y_{l1}Y^{\dagger}_{l1})_{ii} \upsilon^{2}}{m_{N_{1}}} \sim m_{\nu}  $ and $ m_{*}= \dfrac{8\pi \upsilon^{2}}{m^{2}_{N_{1}}}H\vert_{T=m_{N_{1}}}= 1.08 \times 10^{-3}$ eV. Here, $N_1$ is the mass of lightest RHN, whose decay produces the lepton asymmetry.
  The baryon asymmetry is enhanced
 for $(m_{N_{i}}-m_{N_{j}})\sim\Gamma_{i,j} /2$ \cite{Pilaftsis:2003gt}.\\ 
\section{Inverse seesaw model} 
For the sake of thoroughness, we review here the inverse seesaw model \cite{Mohapatra:1986aw, Mohapatra:1986bd}, which was developed and studied in our earlier work \cite{Devi:2021aaz}. We then  discuss how it limits the characteristics of heavy neutrinos.  We use cyclic groups $ Z_{4} $ and $ Z_{5} $  along with  $A_{4}$   group and $ U(1)_{X} $  global symmetry \cite{Devi:2021aaz},. In addition to the SU(2) singlet neutrinos, $ N$, our model also includes a family of three singlet sterile neutrinos, designated as S. The neutrino mass matrix generated from the seesaw model's Lagrangian in the basis  ($ \nu^{c}_{L}, N, S$)  can be expressed as follows \cite{Wyler:1982dd}:\\
\begin{equation}
 \left( M_{\nu} \right)_{iss}=\begin{bmatrix}
0 & m_{D} & 0\\
m^{T}_{D} & 0 & M\\
0 & M^{T} & \mu_{s}
\end{bmatrix} \textrm{.}  
\label{ISS: ISS mass matrix}
\end{equation}\\
\\
Here, $m_D$ is the Dirac neutrino mass, M is the mass matrix of RH Majorana neutrinos, and $ \mu_{s}$ is the mass matrix of the sterile neutrinos. The prerequisite $m_{D}, M \gg \mu_{s}$ must be met in order to implement the inverse seesaw mechanism. The light neutrino mass matrix is doubly suppressed by M and depends on the previous condition to produce it at $\mathcal{O}(eV) $ scale as
\begin{equation}
m_{\nu}= m_{D}(M^{T})^{-1}\mu_{s} M^{-1}m^{T}_{D}\textrm{ .}
\label{ISS:neutrino_matrix_equation}
\end{equation}
\\
The lepton number conservation is broken down in this instance by the  $\mu_{s}$  term of the low-scale seesaw mechanism, as noted in \cite{tHooft:1980xss}. In Table (\ref{tab:vevA4ISS}), where  $\Phi_s$, $\Phi_t$ are triplet scalar fields and  $ \eta $, $\xi$, $\tau$, $\rho$, $\rho^{\prime}$, $\rho^{\prime \prime}$  are singlet scalar fields under  $ A_{4} $ group transformation, we show the particle content of the various fields included in the model. $H$ is the SM Higgs doublet, and $L$ is the SM LH leptonic doublet.
\begin{table}[h]
\begin{center}
\begin{tabular}{|c|cc|ccc|cc|cccccccc|}
\hline
 & L & H & $ e_{R} $ & $ \mu_{R} $ & $ \tau_{R} $ & N & S & $\Phi_T$ & $\Phi_s$ & $\eta$ & $\xi$ & $\tau$ & $\rho$ & $\rho'$ & $\rho^{\prime\prime}$\\
\hline
$A_4$ & 3 & 1 & 1 & $ 1^{\prime\prime} $ & $ 1^{\prime} $ & 3 & 3 & 3 & 3 & 1 & $1'$ & $1''$ & 1 & 1 & 1 \\
\hline 
$Z_4$ & 1 & 1 & i & i  & i & i & 1 & i & -i & -i & -i & -i & i & i & 1 \\
\hline
$Z_{5}$ & 1  & 1 & $ \omega $ & $ \omega $  & $ \omega $ & $ \omega^{2} $ & 1 & $ \omega$ & 1 & 1 & 1 &1 & $ \omega^{2} $ & 1 & 1 \\
\hline
$U(1)_{X}$ & -1 & 0 & -1 & -1 & -1 & -1 & 1 & 0 & -1 & -1 & -1 & -1 & 0 & -4 & -3 \\
\hline
\end{tabular}
\end{center}
\caption{Particle content under $ A_{4}\times Z_{4} \times Z_{5} \times U(1)_{X} $ symmetry for inverse seesaw model, taken from  our work \cite{Devi:2021aaz}}
\label{tab:vevA4ISS}
\end{table}
Applying the $ A_{4} $ product rules now to the aforementioned fields yields the following charged lepton's Lagrangian:\\
\begin{equation} 
\mathcal{L}_{c.l.} \supset  \dfrac{y_{e}}{\Lambda}(\bar{L}\Phi^{\dagger}_{T})H e_{R}+\dfrac{y_{\mu}}{\Lambda}(\bar{L}\Phi^{\dagger}_{T})^{\prime}H \mu_{R}+\dfrac{y_{\tau}}{\Lambda}(\bar{L}\Phi^{\dagger}_{T})^{\prime\prime}H \tau_{R}\textrm{ .}
\label{ISS:revised charged lagrangian modified}
\end{equation}\\
In matrix form, Eq. (\ref{ISS:revised charged lagrangian modified}) can be written as:
\begin{equation}
 M_{c.l.}=\dfrac{\upsilon^{\dagger}_{t}\upsilon_{h}}{\Lambda}\begin{bmatrix}
y_{1} & 0 & 0\\
0 & y_{2} & 0\\
0 & 0 & y_{3}
\end{bmatrix} \textrm{,}
\label{Charged lepton mass matrix}
\end{equation}\\
\\
Here, $ y_{e} $, $ y_{\mu}$ and $ y_{\tau} $  are  the coupling constants, the normal cut-off scale of the theory is denoted as $\Lambda$, and the VEVs of the standard model Higgs are taken as $\langle h \rangle= \upsilon_{h} $,  and  $ \langle \Phi_{t} \rangle= \upsilon_{t} $. After using the $ A_{4} $  product rules, which are  $1'\times 1'=1''$, $1'\times 1''=1, 1''\times 1''= 1'$ and $3\times 3=1+1'+1''+3_A+3_S$ \cite{Altarelli:2010gt},  we may move on to writing the neutrino sector Lagrangian which can be expressed as
\begin{equation} 
\mathcal{L}_{\rm Y} \supset  Y_D \frac{\bar{L} \tilde{H} N \rho^{\dagger}}{\Lambda} + Y_M N S \rho^{\dagger} + Y_{\mu} S S [\frac{\rho^{\prime} \rho^{\prime\prime^{\dagger}}(\Phi_s + \eta + \xi + \tau) }{\Lambda^2}  ]+ h.c. \textrm{.}
\label{ISS:revised lagrangian}
\end{equation}
From Eqn. (\ref{ISS:revised lagrangian}), we get various mass matrices as  \cite{Devi:2021aaz}:
\begin{equation}
 M_{D}=\dfrac{Y_{D}\upsilon_{h}\upsilon^{\dagger}_{\rho}}{\Lambda}\left(\begin{matrix}
1 & 0 & 0\\
0 & 0 & 1\\
0 & 1 & 0
\end{matrix}\right) \Rightarrow Y_{D} =  M_{D} \dfrac{\Lambda}{\upsilon_{h}\upsilon^{\dagger}_{\rho}}\left(\begin{matrix}
1 & 0 & 0\\
0 & 0 & 1\\
0 & 1 & 0
\end{matrix}\right) \textrm{, }
\label{ISS: MD}
\end{equation}
\begin{equation}
  M=Y_{M}\upsilon^{\dagger}_{\rho}\left( \begin{matrix}
1 & 0 & 0\\
0 & 0 & 1\\
0 & 1 & 0
\end{matrix}\right) \textrm{, }
\label{ISS:M}
\end{equation}
\begin{equation}
 \textrm{and, } \mu_{s}=\dfrac{Y_{\mu_{s}}\upsilon_{\rho^{\prime}}\upsilon^{\dagger}_{\rho_{\prime\prime}}}{\Lambda^{2}}\left(
\begin{array}{ccc}
v_{\Omega }+2 v_s \phi _a & v_{\xi }-v_s \phi _c & v_{\tau }-v_s \phi _b \\
 v_{\xi }-v_s \phi _c & v_{\tau }+2 v_s \phi _b & v_{\Omega }-v_s \phi _a \\
 v_{\tau }-v_s \phi _b & v_{\Omega }-v_s \phi _a & v_{\xi }+2 v_s \phi _c \\
\end{array}
\right) \textrm{.}
\label{ISS:Mu}
\end{equation}
\begin{equation}
 \Rightarrow m_{\nu}=F_{1}\left(
\begin{array}{ccc}
 v_{\Omega }+2 v_s \phi _a & v_{\xi }-v_s \phi _c & v_{\tau }-v_s \phi _b \\
 v_{\xi }-v_s \phi _c & v_{\tau }+2 v_s \phi _b & v_{\Omega }-v_s \phi _a \\
 v_{\tau }-v_s \phi _b & v_{\Omega }-v_s \phi _a & v_{\xi }+2 v_s \phi _c \\
\end{array}
\right)\textrm{,}
\label{ISS:final matrix with vev}
\end{equation}
where, $ F_{1}= \dfrac{Y^{2}_{D}Y_{\mu_{s}}}{Y^{2}_{M}}[\dfrac{v^{2}_{h}v_{\rho^{\prime}}v^{\dagger}_{\rho^{\prime \prime}}}{\Lambda^{4}}]$.
Here, $ Y_D $, $ Y_M $, $ Y_{\mu_{s}} $ are the dimensionless coupling constants which are usually complex. The non-zero VEVs of scalars can be represented as:  $ \langle H \rangle= v_{h} $,     $ \langle \Omega \rangle= v_{\Omega} $, 
$ \langle\rho \rangle= v_{\rho} $, $ \langle\rho^{\prime} \rangle= v_{\rho^{\prime}} $, $ \langle\rho^{\prime \prime} \rangle= v_{\rho^{\prime \prime}} $, $ \langle \xi \rangle= v_{\xi} $,  $ \langle \tau \rangle= v_{\tau} $,  $ \langle\Phi_{S} \rangle= v_{s}(\Phi_{a},\Phi_{b},\Phi_{c}) $.\\
\\
Thus from Eqns. (\ref{ISS:M}), (\ref{ISS:Mu})
we get the mass of degenerate RH Majorana neutrinos $ M_{R}=M \pm \mu_{s}/2 $, i.e.
\begin{equation}
M_{R_{\pm}}=Y_{M}\upsilon^{\dagger}_{\rho}\left( \begin{matrix}
1 & 0 & 0\\
0 & 0 & 1\\
0 & 1 & 0
\end{matrix}\right) \pm \dfrac{K}{2}\left(
\begin{array}{ccc}
A +2  \phi _a & C- \phi _c & B-\phi _b \\
 C- \phi _c & B+2 \phi _b & A- \phi _a \\
 B- \phi _b & A- \phi _a & C+2 \phi _c \\
\end{array}
\right)
\end{equation}
where each non-zero element is M $ \sim \mathcal{O}(10^{12}) $ eV, $ K = \dfrac{Y_{\mu_{s}}v_s\upsilon_{\rho^{\prime}}\upsilon^{\dagger}_{\rho_{\prime\prime}}}{\Lambda^{2}}\sim \mathcal{O}(1)$ eV. We scan the parameters $A,B,C  \sim (1-9)\times 10^{3}  $ where $A=v_{\Omega }/v_s$, $C=v_{\xi }/v_s$, $B=v_{\tau }/v_s$ and are dimensionless parameters.
\section{Numerical analysis}
We compute the BAU in the  low-scale  resonant  leptogenesis scenario as described in section \ref{sec:RL2}.   On comparing Eqn. (6) with the light neutrino mass matrix obtained using the global best fit values with a parametrisation  of $U_{PMNS}$ matrix, we get a system of equations for the triplet flavon components.  Then we scan the unknown parameters space (as input) in a viable range.The information obtained by numerically solving this system of equations can be used to compute BAU in our model, using equations (1-3).  We have done our computation up to the tolerance $< 10^{-5}$.\\
\\
As the scale of flavor symmetry breaking is still not known, therefore, in Eqn. (\ref{ISS: MD}), we take different values of $ \dfrac{\upsilon^{\dagger}_{\rho}}{\Lambda} $ for different cases, and compute $ \epsilon $ and $ \eta_{B} $ accordingly. Mass matrix $ M_{D} $ is parametrised by fitting the light-neutrino data in the Extended Casas Ibarra parametrisation \cite{Dolan:2018qpy, Dolan:2018yqy}
\begin{equation}
\Rightarrow M_{D}=Um^{1/2}R\mu_{s}^{-1/2}M^{T} 
\label{casas}
\end{equation}
here, R is taken as complex orthogonal matrix \cite{Dolan:2018yqy} $ RR^{T}=\mathbf{1}_{3\times 3} $ with $ \theta_{n}=Re(\theta_{n})+i Im(\theta_{n}) $
\begin{equation}
R(\theta_{1},\theta_{2},\theta_{3})=R(\theta_{1})R(\theta_{2})R(\theta_{3})
\end{equation}
\begin{equation}
R(\theta_{1})=\left( \begin{matrix}
c_{1} & -s_{1} & 0\\
s_{1} & c_{1} & 0\\
0 & 0 & 1
\end{matrix}\right), R(\theta_{2})=\left( \begin{matrix}
c_{2} & 0 & -s_{2}\\
0 & 1 & 0\\
s_{2} & 0 & c_{2}
\end{matrix}\right), R(\theta_{3})=\left( \begin{matrix}
1 & 0 & 0\\
0 & c_{3} & -s_{3}\\
0 & s_{3} & c_{3}
\end{matrix}\right)
\end{equation}
where $ c_{1} =cos\theta_{1} $,  $ c_{2} =cos\theta_{2} $,  $ c_{3} =cos\theta_{3} $ and  $ s_{1} =sin\theta_{1} $, $ s_{2} =sin\theta_{2} $, $ s_{3} =sin\theta_{3} $. The values of the real and imaginary components of $ \theta_{1} $, $ \theta_{2} $ and $ \theta_{3} $ (randomly chosen) are fixed as:
\begin{equation}
Re[\theta_{1}]=\dfrac{\pi}{5}, Re[\theta_{2}]=\dfrac{5\pi}{6},
Re[\theta_{3}]=\dfrac{4\pi}{7},  
Im[\theta_{1}]=Im[\theta_{3}]=0, Im[\theta_{2}]=10^{-2}.  
\end{equation}\\
where there is no specific significance to the chosen value of the above angles. The resonant leptogenesis parameters are not sensitive to these values and hence any value of above angles can be used.
\section{Results and Discussions} 
By examining whether the deviation from thermal equilibrium during baryogenesis took place through a freeze-out or free-in scenario, one may categorise the type of leptogenesis one has to deal with. Typically, heavy neutrinos with degenerate Majorana masses in the freeze-out framework are associated with resonant leptogenesis. The masses of  heavy Majorana neutrinos,  coupling constants and mass splitting between degenerate Majorana masses decide the magnitude of the BAU. In what follows,  we discuss some parameters and issues that affect the value of BAU.
\subsection{Prerequisites and assessment of parameter analysis}
We solve a set of flavon equations of triplet scalar $ \Phi_{s} $ that is obtained after comparing Eqn. (\ref{ISS:neutrino_matrix_equation}) derived from our model with the light neutrino mass matrix obtained after standard parametrisation $ U_{PMNS}m_{\nu_{diag}}U^{T}_{PMNS} $. In these flavon equations, whose solutions provide us with the values of the unknown neutrino oscillation parameters, we do a parameter scan for the known sets of neutrino oscillation parameters involved. We examined \cite{Devi:2021aaz} the potential parameter space for the normal and inverted hierarchies within the 3$\sigma$ range of the known parameters, while taking into account various VEV alignments of the triplet scalar flavon $ \Phi_{s}$. It was shown in  \cite{Devi:2021aaz} that only six VEVs are valid in particular mass hierarchies, namely (0,1,1)/(0,-1,1) in NH, (-1,1,1)/(1,-1,-1) in NH, and (0,1,-1)/(0,-1,1) in IH. To compute the baryon asymmetry of the universe while taking into account the 23  elements of mass matrix in equation (14) as $\sim 1$ TeV, we use the data of the oscillation parameters for each of the permissible VEV alignments for the purpose.  From equations (1-4), the connection between leptogenesis and heavy Majorana neutrinos  of about 1 TeV mass is evident, and we can use this to trace the evolution of the baryon asymmetry.\\
\\
\\
Since so far, there is no evidence in support of a particular value of scale of flavour symmetry breaking, in the models one can take a viable range of this scale and investigate its consequences on various physical observables. In our earlier works \cite{Devi:2021aaz,Bora:2022jdp,Devi:2022scm}, we discussed about this. In this work therefore, we compute lepton asymmetry $ \epsilon $ and and BAU $ \eta_B $ for four different values of $ \dfrac{\upsilon^{\dagger}_{\rho}}{\Lambda} $, for all the three allowed VEVs (as shown in our earlier work  \cite{Devi:2021aaz}) of the triplet scalar flavon, $ \Phi_{s} $, in Table (\ref{Tab:lept}).
\begin{table}[ht]
\centering
\subfloat[VEV (-1,1,1) in NH]{
\begin{tabular}{| l | c | c |}
\hline 
\multirow{2}{*}{ $ \dfrac{\upsilon^{\dagger}_{\rho}}{\Lambda} $ } & \multicolumn{2}{c|}{VEV (-1,1,1) in NH}  \\ 
\cline{2-3}
  & Range of  $ \epsilon $ & Range of $ \eta_{B} $   \\ 
\hline
0.1 & $ {-4.40209\times 10^{-6}\rightarrow 0.000151268}$ & $  {2.06889\times 10^{-13}\rightarrow 1.84782\times 10^{-9}}$ \\
0.3 & $ {-0.0000396188\rightarrow 0.00136141} $ & $ {1.862\times10^{-12}\rightarrow 1.66304\times10^{-8}} $ \\
0.5 & $ {-0.000110052\rightarrow 0.0037817} $ & $ {5.17222\times 10^{-12}\rightarrow 4.61955\times 10^{-8}} $    \\
0.7 & $ {-0.000215703\rightarrow 0.00741213} $ & $ {1.01375\times 10^{-11}\rightarrow 9.05433\times 10^{-8}} $\\
\hline 
\end{tabular} \label{tab:a}}
\qquad
\subfloat[VEV (0,1,1) in NH]{
\begin{tabular}{| l | c | c |}
\hline 
\multirow{2}{*}{ $ \dfrac{\upsilon^{\dagger}_{\rho}}{\Lambda} $ } & \multicolumn{2}{c|}{VEV (0,1,1) in NH}  \\ 
\cline{2-3}
  & Range of $ \epsilon $ & Range of $ \eta_{B} $   \\ 
\hline
0.1 & ${-1.65326\times 10^{-6}\rightarrow 7.87137\times 10^{-6}} $ & $ {8.80658\times 10^{-19}\rightarrow 6.61007\times 10^{-11}} $ \\
0.3 & ${-0.0000148793 \rightarrow 0.0000708423} $ & $ {7.92592\times 10^{-18}\rightarrow 5.94906\times 10^{-10}} $ \\
0.5 & $ {-0.0000413314\rightarrow 0.000196784}$ & ${2.20165\times 10^{-17}\rightarrow 1.65252\times 10^{-9}}  $ \\
0.7 & $ {-0.0000810096\rightarrow 0.000385697}$ & $ {4.31523\times 10^{-17}\rightarrow 3.23894\times 10^{-9}} $ \\
\hline  
\end{tabular} \label{tab:b}}
\qquad
\subfloat[VEV (0,1,-1) in IH]{
\begin{tabular}{| l | c | c |}
\hline 
\multirow{2}{*}{ $ \dfrac{\upsilon^{\dagger}_{\rho}}{\Lambda} $ } & \multicolumn{2}{c|}{VEV (0,1,-1) in IH}  \\ 
\cline{2-3}
  & Range of $ \epsilon $ & Range of $ \eta_{B} $   \\ 
\hline
0.1 & $ {-3.13823\times 10^{-7}\rightarrow 0.0000242336}$ & $ {3.22169\times 10^{-15} \rightarrow 2.08037\times 10^{-10}} $ \\
0.3 & ${-2.82441\times 10^{-6}\rightarrow 0.000218103} $ & $ {2.89952\times 10^{-14}\rightarrow 1.87234\times 10^{-9}} $ \\
0.5 & ${-7.84558 \times 10^{-6}\rightarrow 0.000605841} $ & ${8.05422 \times 10^{-14}\rightarrow 5.20093 \times 10^{-9}}  $ \\
0.7 & ${-0.0000153773\rightarrow 0.00118745} $ & $ {1.57863\times 10^{-13}\rightarrow 1.01938\times  10^{-8}} $ \\
\hline 
\end{tabular} \label{tab:c}}
\caption{Ranges of parameters $ \epsilon $ and $ \eta_{B} $ for VEV (-1,1,1) in NH, VEV (0,1,1) in NH and VEV (0,1,-1) in IH are shown in Tables (\ref{tab:a}), (\ref{tab:b}) and (\ref{tab:c}) respectively}
  \label{Tab:lept}
\end{table}          
\\
\\
We have also obtained the correlation plots ($\eta_{B}\times 10^{10}$ vs $ Log [\kappa ]$), ($\eta_{B}\times 10^{10}  $ vs $ Log [\epsilon ] $), ($\eta_{B}\times 10^{10}$ vs $ m_{3} $) and ($ \eta_{B}\times 10^{10}$ vs $ (Sin \delta)^{2}$),  for the three allowed VEV alignments of the triplet scalar flavon, and are shown in  Figs. (\ref{Fig-111 A}- \ref{01-1 D}), where (\textcolor{cyan}{\textbf{- -}}) dashed and (\textcolor{purple}{\textbf{---}}) bold lines represents the limits given by $ 5.8 \leq \eta_{B}\times 10^{-10} \leq 6.5 $\cite{Riemer-Sorensen:2017vxj, Cooke:2017cwo} and $ \eta^{obs}_{B}= (6.5 \pm 0.08)\times 10^{-10}$, \cite{Planck:2018vyg} respectively.\\
\subsection{Impact of Coupling constant and flavon VEV alignment}
One of the the primary factors affecting the production and decay rates of RH neutrinos is the magnitude of the relevant Yukawa couplings. Small Yukawa couplings result in lesser production of RH neutrinos and smaller decay rates, which might result in a BAU value that is lower than the observed value as is evident from Eqn. (\ref{epsilon}).  Fortunately, this limit can be circumvented for special choices of parameters such as flavon VEVs $ \upsilon^{\dagger}_{\rho} $ and the cut-off scale of the theory related to seesaw mechanism $ \Lambda$, for which the size of the Yukawa couplings can be sufficient enough,  to produce baryogenesis  via Sphaleron processes within its global fit parameter space. From Eqn. (\ref{ISS: MD}), we can infer that there is a decrease in the production of right-handed Majorana neutrinos as the ratio of $\dfrac{\upsilon^{\dagger}_{\rho}}{\Lambda} $ increases. This can be justified because the Yukawa couplings  gradually decrease with the increasing value of $\dfrac{\upsilon^{\dagger}_{\rho}}{\Lambda}$. The computed information on dynamics of the theory, i.e. VEV of triplet flavon field and values of the Yukawa couplings, for which the light neutrino mass matrix of equation (13) lies in its experimentally allowed $3\sigma$ range, are shown in Table (\ref{tab:yuk}).
\begin{table}[h]
        \centering
        \begin{tabular}{| l | c | c | c |}
\hline 
\multirow{3}{*}{ $ \upsilon^{\dagger}_{\rho}/{\Lambda} $ } & \multicolumn{3}{c|}{Range of Yukawa Coupling}  \\ 
\cline{2-4}
  & VEV (-1,1,1)in  NH & VEV (0,-1,1)in  IH  &  VEV (0,1,1)in  NH \\ 
\hline
0.1 & $ { 0.0485142\rightarrow 0.431652 } $ & $ {0.117593 \rightarrow 0.549414 } $ & $ {0.000250288 \rightarrow  0.518041} $ \\
0.3 & $ { 0.0161714\rightarrow 0.143884 } $ & $ { 0.0391976\rightarrow 0.0183138 } $ & $ { 0.0000834292\rightarrow 0.17268 } $\\
0.5 & $ { 0.00970285\rightarrow 0.0863305 } $ & $ { 0.023185\rightarrow 0.109883 } $ & $ { 0.000000575\rightarrow 0.103608 } $\\
0.7 & $ { 0.00693061\rightarrow 0.0616646 } $ & $ {0.016799 \rightarrow 0.0784878 } $ & $ {0.0000357554 \rightarrow 0.0740058 } $\\
\hline 
\end{tabular}
      \caption{Ranges of Yukawa couplings for different VEV alignments of the triplet  flavon  field}
        \label{tab:yuk}
             \end{table}\\             
\subsection{Effect of Washout on Leptogenesis}
In addition to the mass splittings and decay width of the RH neutrinos, the amount of BAU generated by low-scale leptogenesis also depends on how quickly any asymmetry is eluded through washout processes.
\begin{figure}[h]
\centering
\subfloat[]{\includegraphics[width=0.48\textwidth]{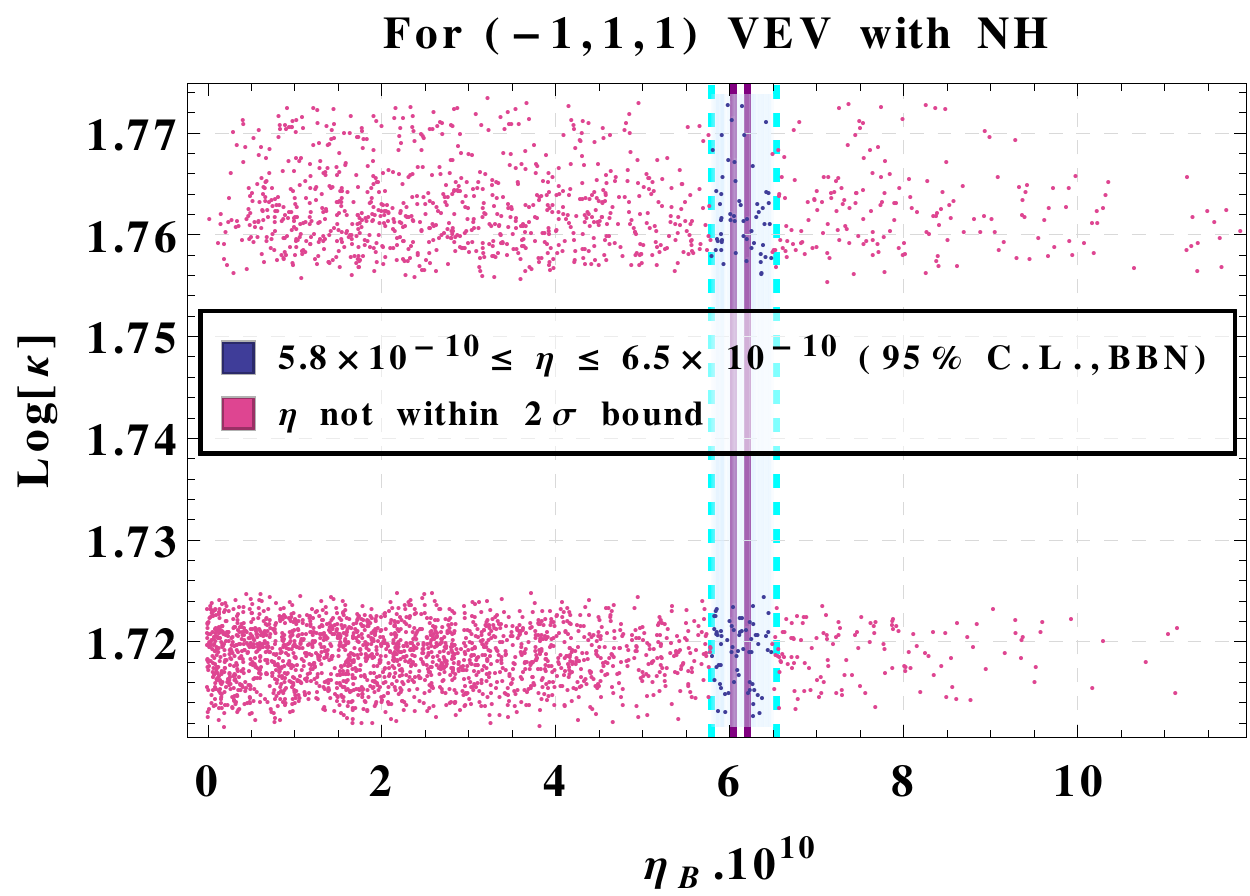}\label{fig1a}}
\qquad
\subfloat[]{\includegraphics[width=0.47\textwidth]{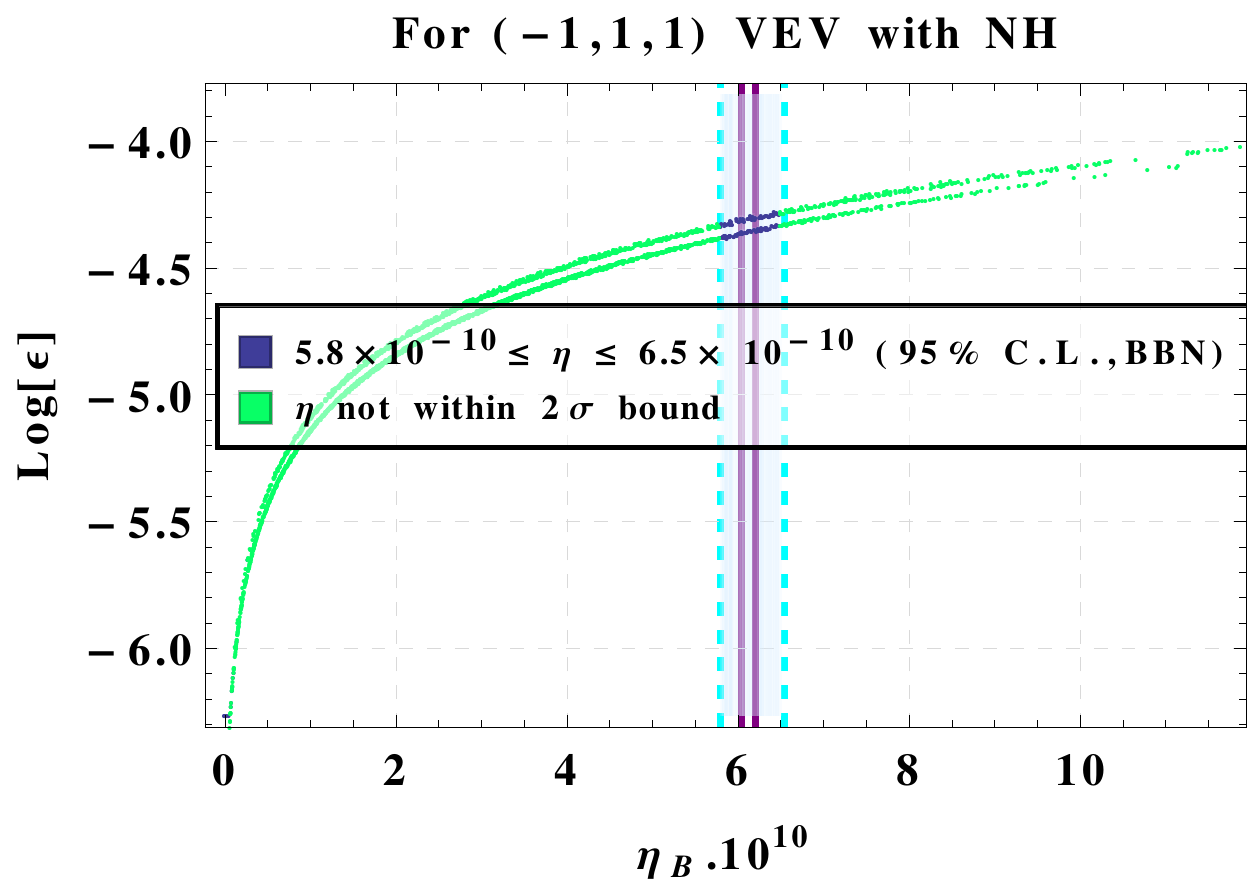}\label{fig1b}}
\qquad
\subfloat[]{\includegraphics[width=0.48\textwidth]{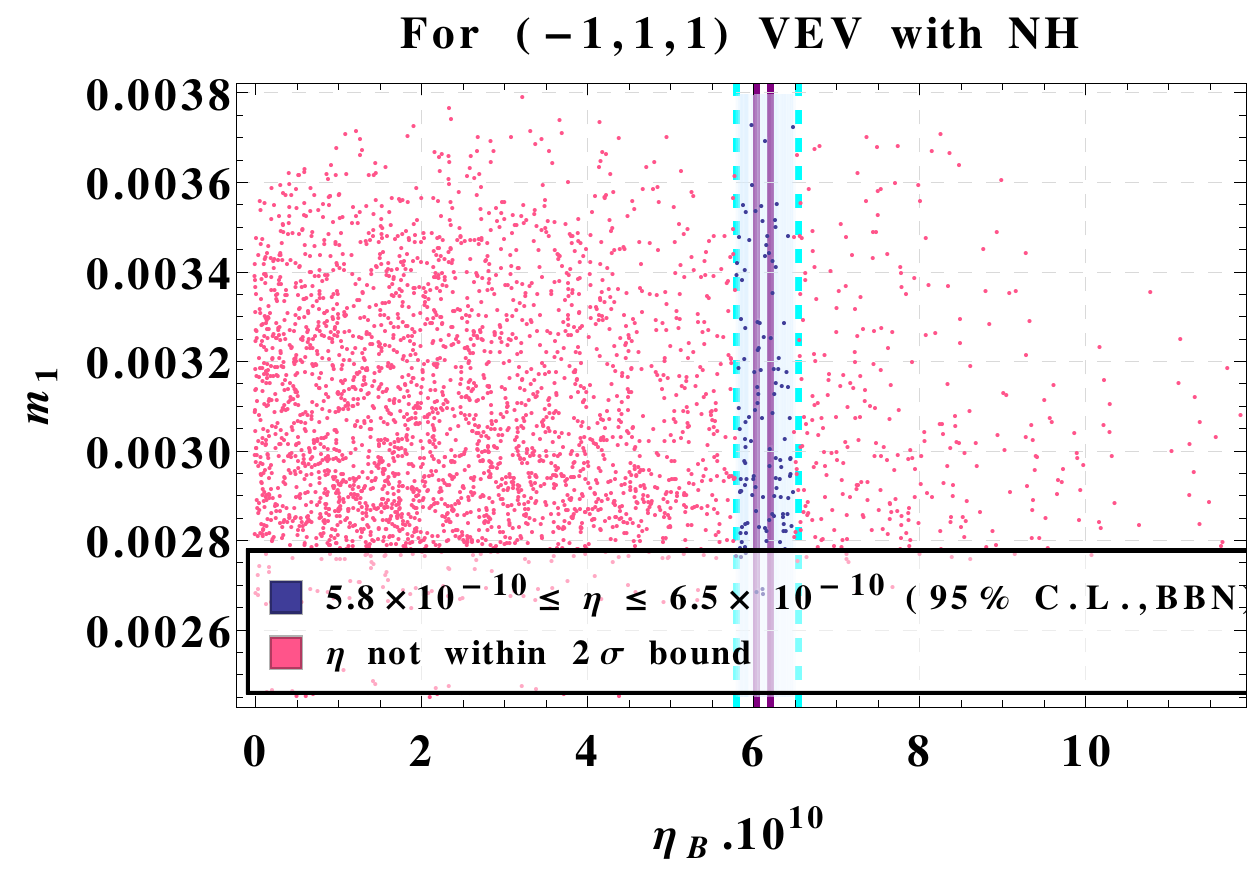}\label{fig1c}}
\qquad
\subfloat[]{\includegraphics[width=0.47\textwidth]{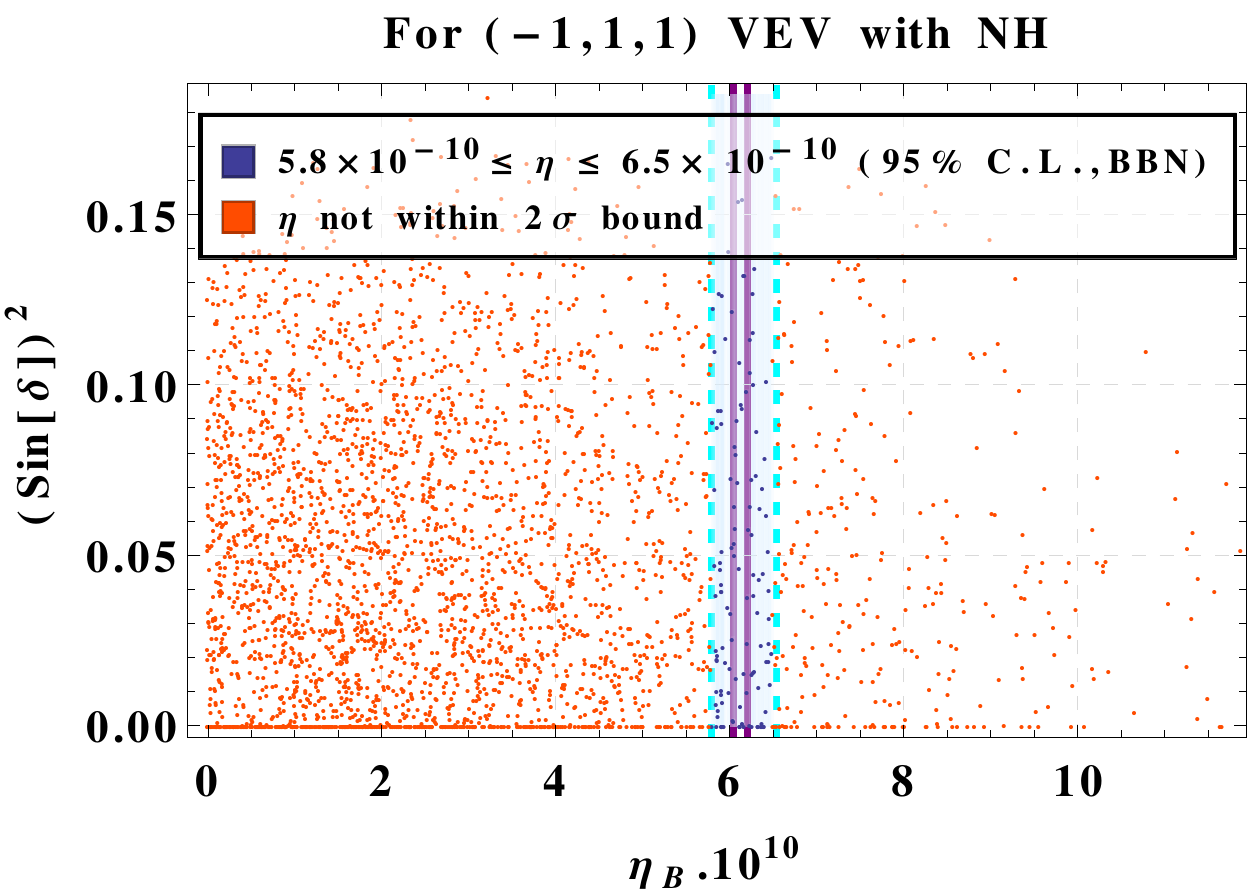}\label{fig1d}}
\caption{Different correlation plots between neutrino oscillation parameters and resonant leptogenesis parameters, as computed in our work. The plots (\ref{fig1a}), (\ref{fig1b}), (\ref{fig1c})  and (\ref{fig1d}) show the correlation between (i) $ \eta_{B}.10^{10} $ vs Log[$ \kappa $], (ii) $ \eta_{B}.10^{10} $ vs Log[$ \epsilon $], (iii) $ \eta_{B}.10^{10} $ vs $ m_{1} $ and (iv) $ \eta_{B}.10^{10} $ vs $(Sin \delta )^{2}$  for (-1,1,1) NH with $\dfrac{\upsilon^{\dagger}_{\rho}}{\Lambda}\sim 0.1$ respectively.  The (\textcolor{cyan}{\textbf{- -}}) dashed and (\textcolor{purple}{\textbf{---}}) bold lines represents the limits given by $ 5.8 \leq \eta_{B}\times 10^{-10} \leq 6.5 $\cite{Riemer-Sorensen:2017vxj, Cooke:2017cwo} and $ \eta^{obs}_{B}= (6.5 \pm 0.08)\times 10^{-10}$, \cite{Planck:2018vyg} respectively.}
\label{Fig-111 A}
\end{figure}
\begin{figure}[h]
\centering
\subfloat[]{\includegraphics[width=0.48\textwidth]{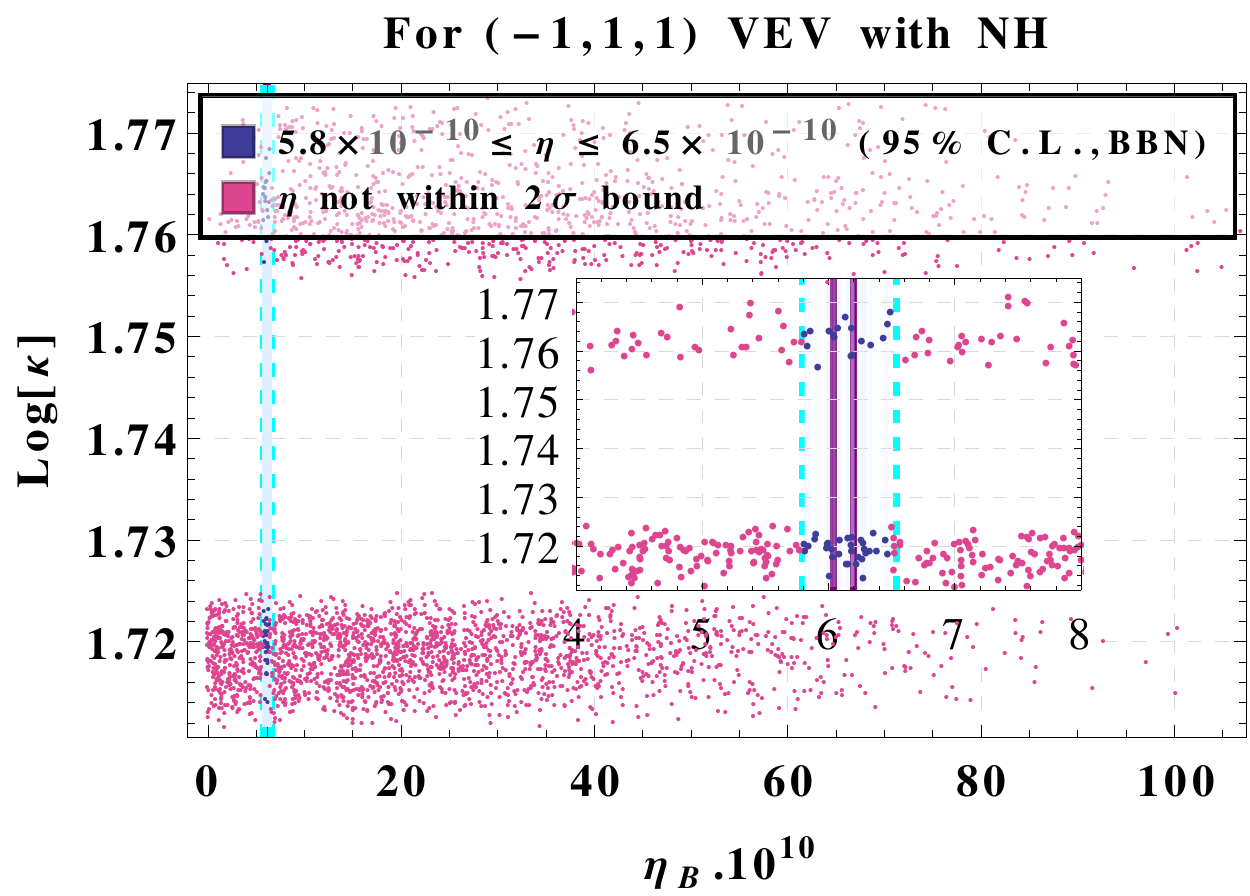}\label{fig2a}}
\qquad
\subfloat[]{\includegraphics[width=0.47\textwidth]{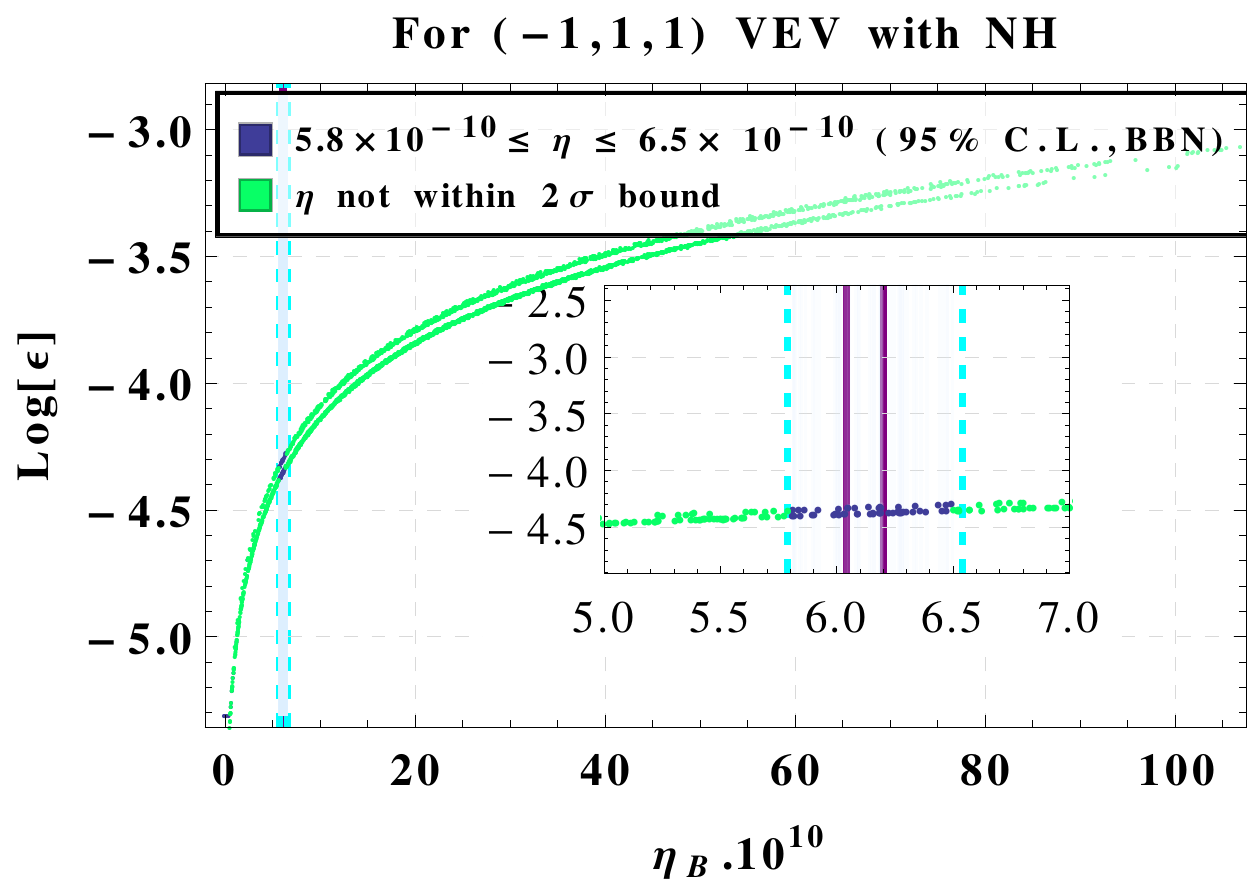}\label{fig2b}}
\qquad
\subfloat[]{\includegraphics[width=0.48\textwidth]{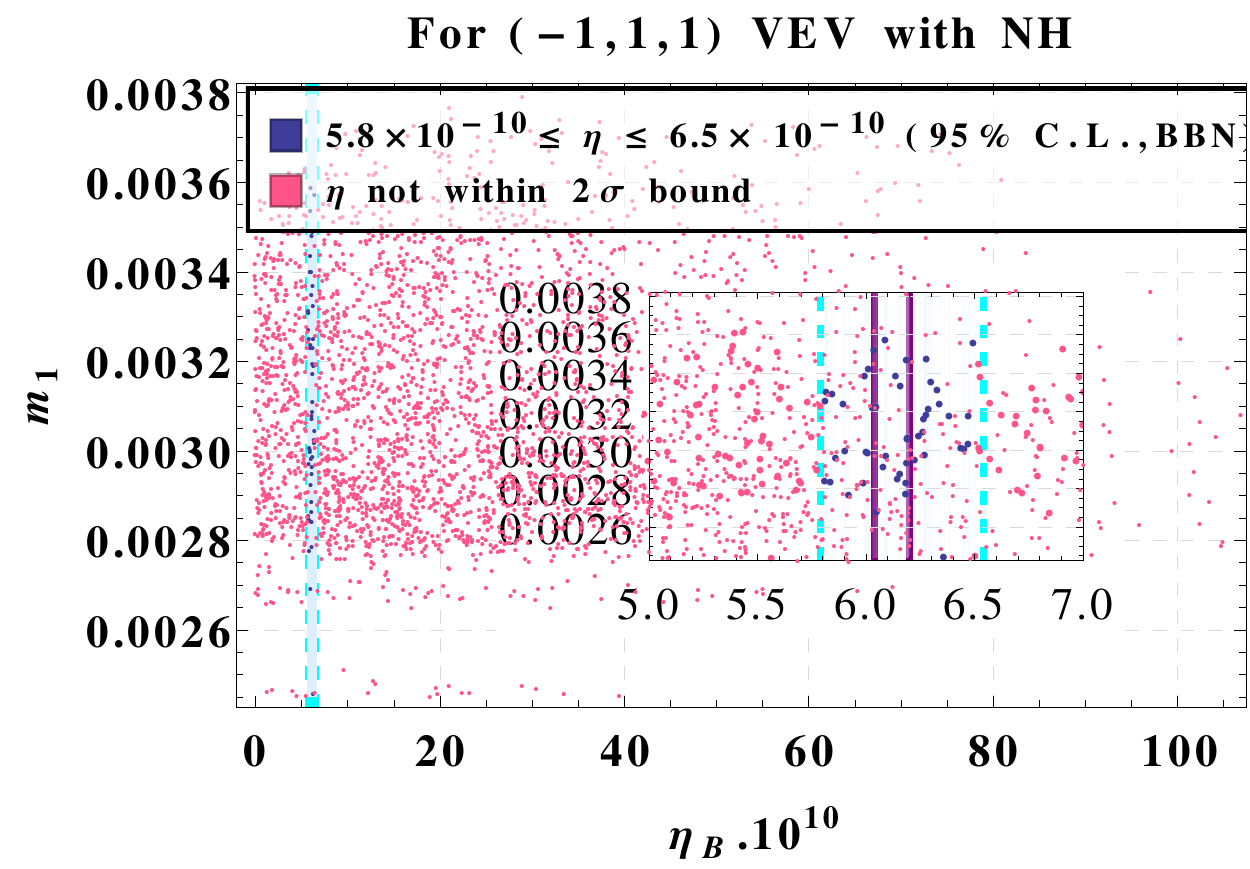}\label{fig2c}}
\qquad
\subfloat[]{\includegraphics[width=0.47\textwidth]{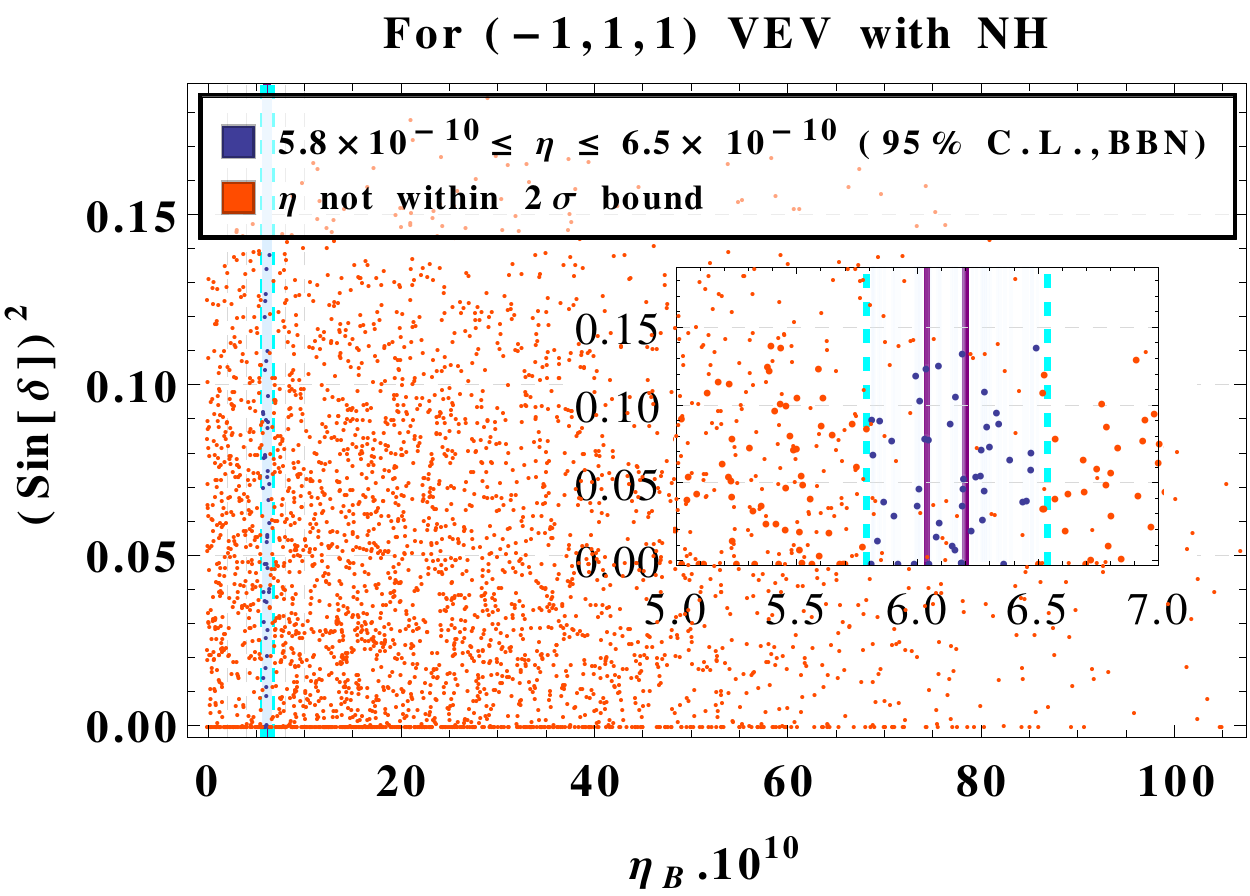}\label{fig2d}}
\caption{The plots (\ref{fig2a}), (\ref{fig2b}), (\ref{fig2c})  and (\ref{fig2d}) show the correlation between (i) $ \eta_{B}.10^{10} $ vs Log[$ \kappa $], (ii) $ \eta_{B}.10^{10} $ vs Log[$ \epsilon $], (iii) $ \eta_{B}.10^{10} $ vs $ m_{1} $ and (iv) $ \eta_{B}.10^{10} $ vs $(Sin \delta )^{2}$  for (-1,1,1) NH with $\dfrac{\upsilon^{\dagger}_{\rho}}{\Lambda}\sim 0.3$ respectively.}
\label{Fig-111 B}
\end{figure}
\begin{figure}[h]
\centering
\subfloat[]{\includegraphics[width=0.48\textwidth]{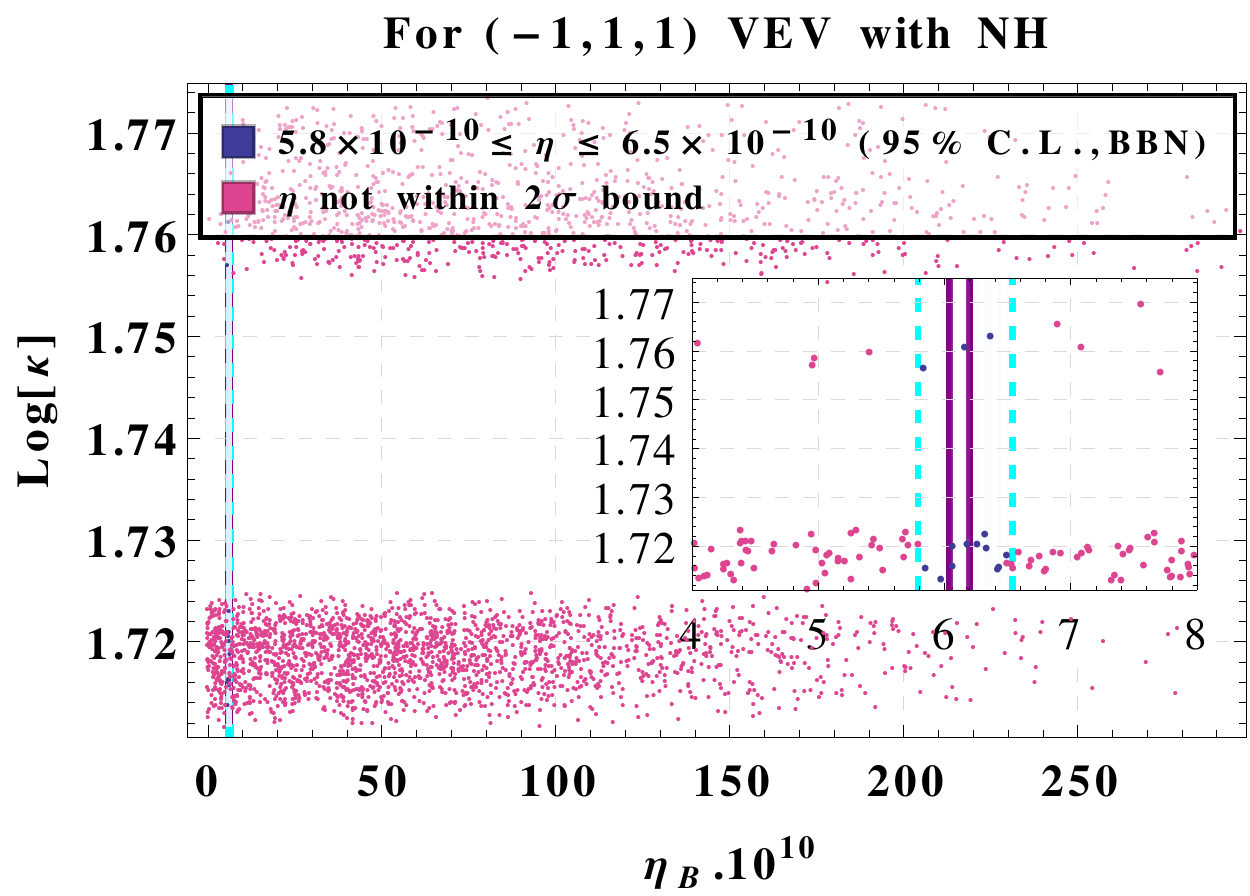}\label{fig3a}}
\qquad
\subfloat[]{\includegraphics[width=0.47\textwidth]{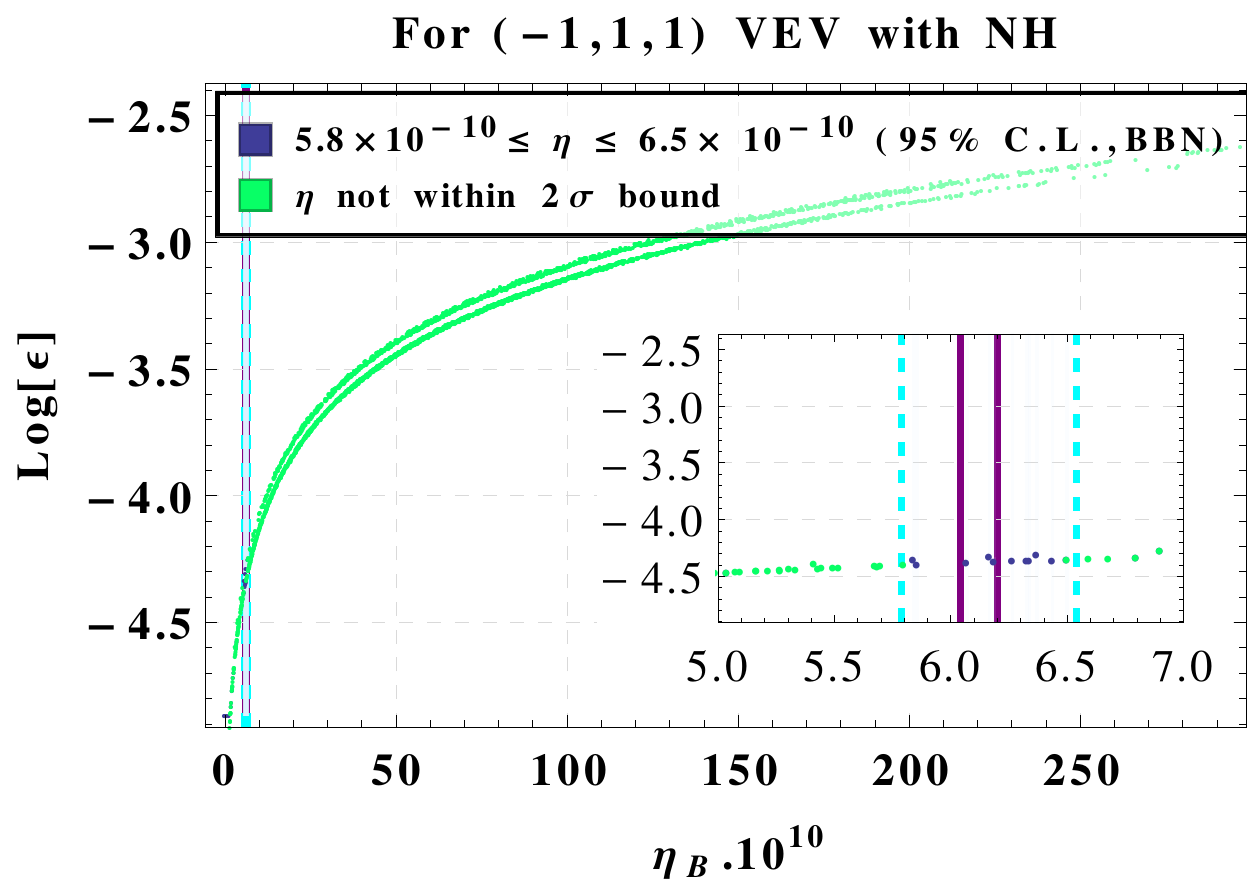}\label{fig3b}}
\qquad
\subfloat[]{\includegraphics[width=0.48\textwidth]{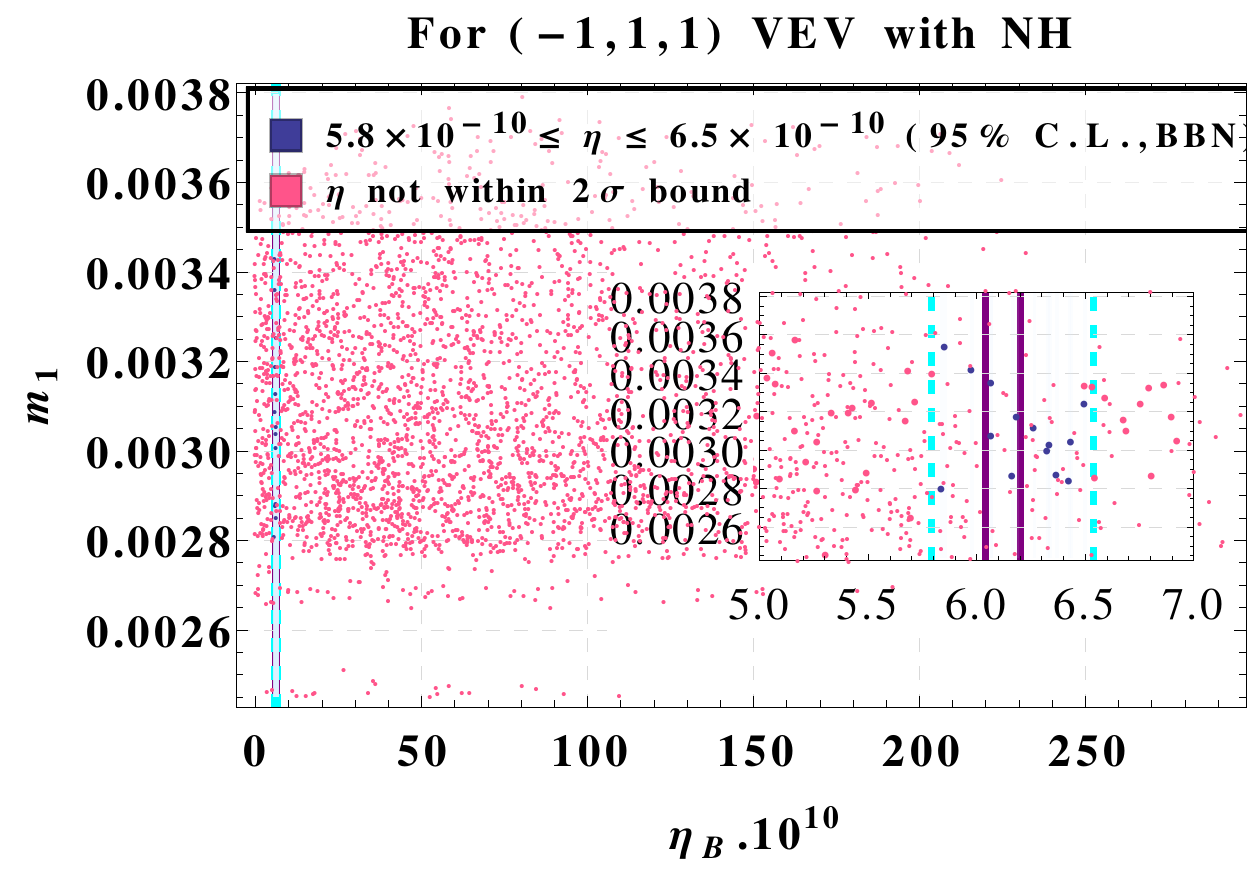}\label{fig3c}}
\qquad
\subfloat[]{\includegraphics[width=0.47\textwidth]{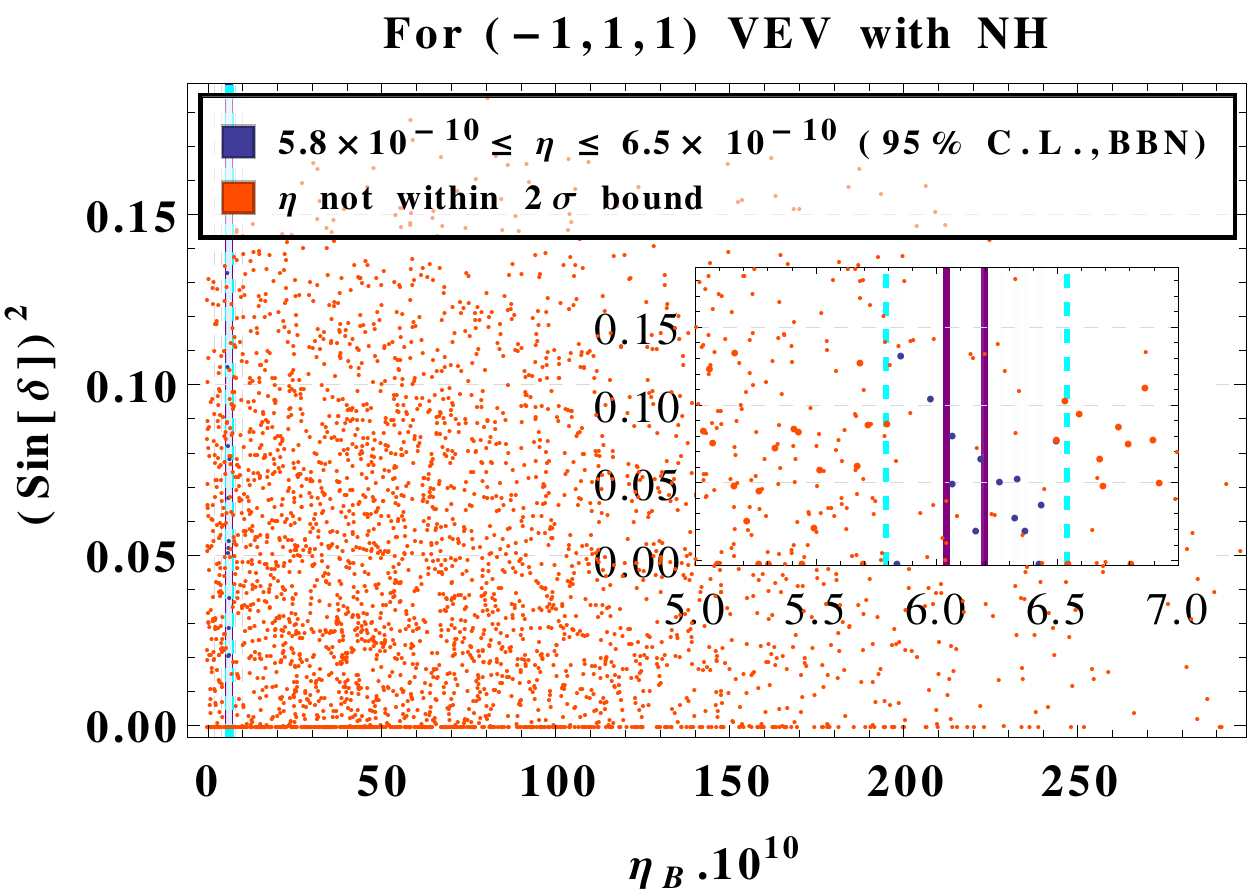}\label{fig3d}}
\caption{The plots (\ref{fig3a}), (\ref{fig3b}), (\ref{fig3c})  and (\ref{fig3d})  show the correlation between (i) $ \eta_{B}.10^{10} $ vs Log[$ \kappa $], (ii) $ \eta_{B}.10^{10} $ vs Log[$ \epsilon $], (iii) $ \eta_{B}.10^{10} $ vs $ m_{1} $ and (iv) $ \eta_{B}.10^{10} $ vs $(Sin \delta )^{2}$  for (-1,1,1) NH with $\dfrac{\upsilon^{\dagger}_{\rho}}{\Lambda}\sim 0.5$ respectively.}
\label{Fig-111 C}
\end{figure}
\begin{figure}[h]
\centering
\subfloat[]{\includegraphics[width=0.48\textwidth]{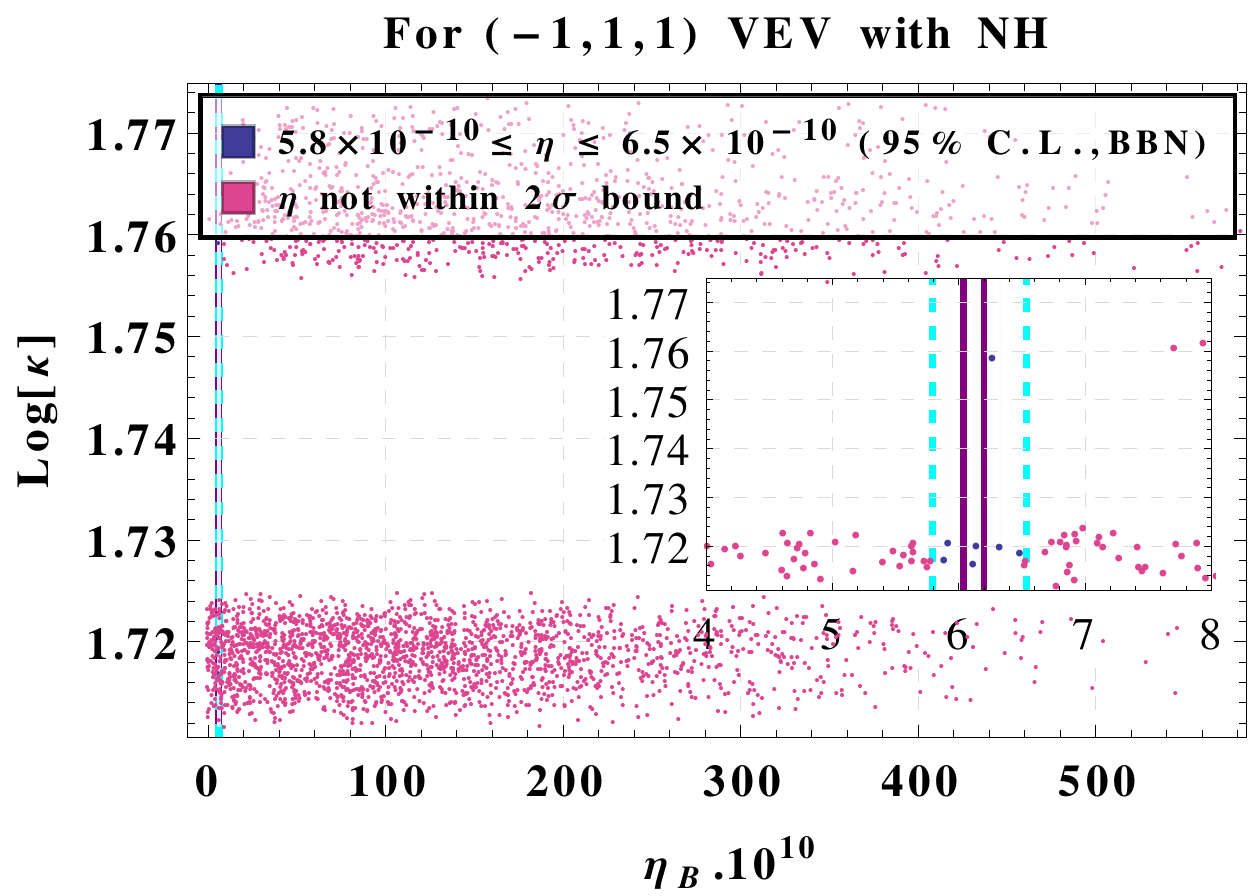}\label{fig4a}}
\qquad
\subfloat[]{\includegraphics[width=0.47\textwidth]{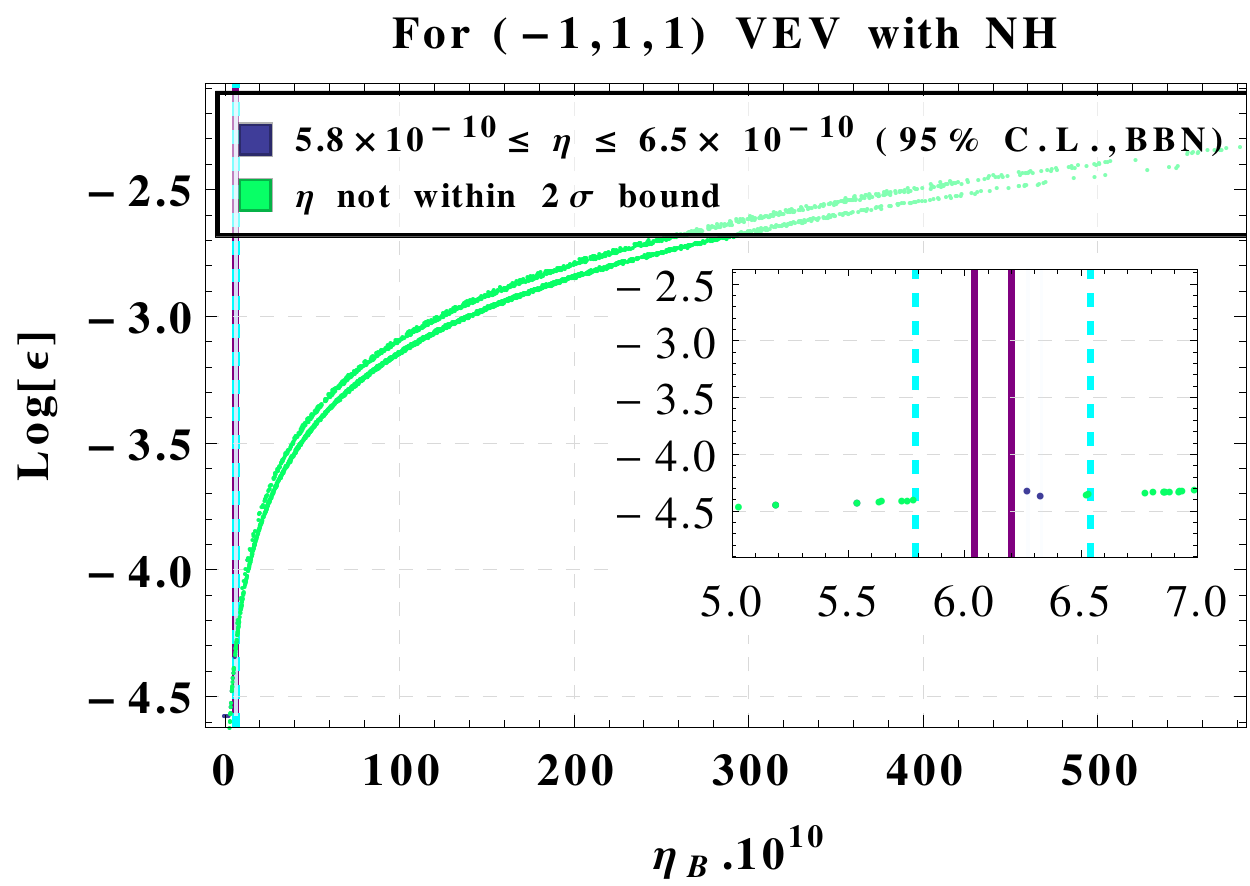}\label{fig4b}}
\qquad
\subfloat[]{\includegraphics[width=0.48\textwidth]{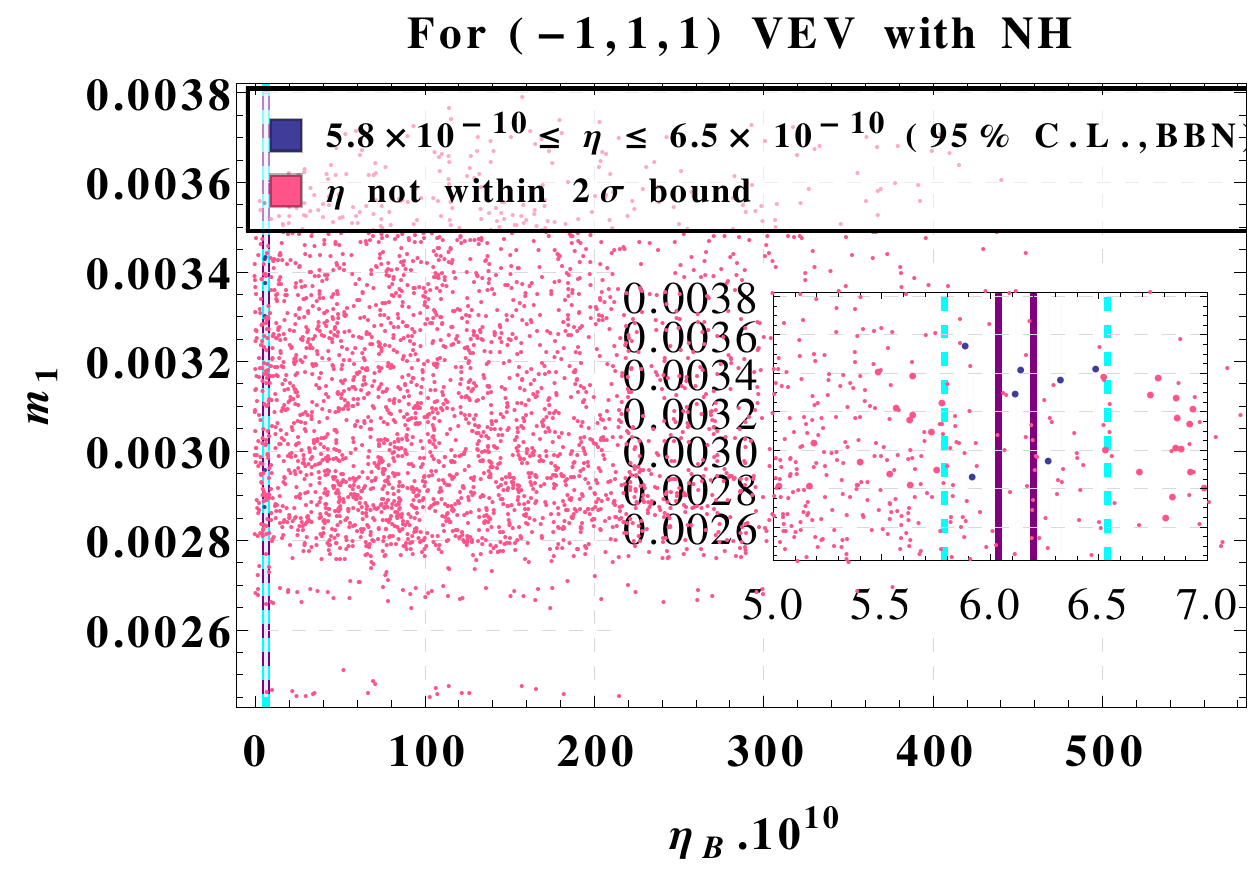}\label{fig4c}}
\qquad
\subfloat[]{\includegraphics[width=0.47\textwidth]{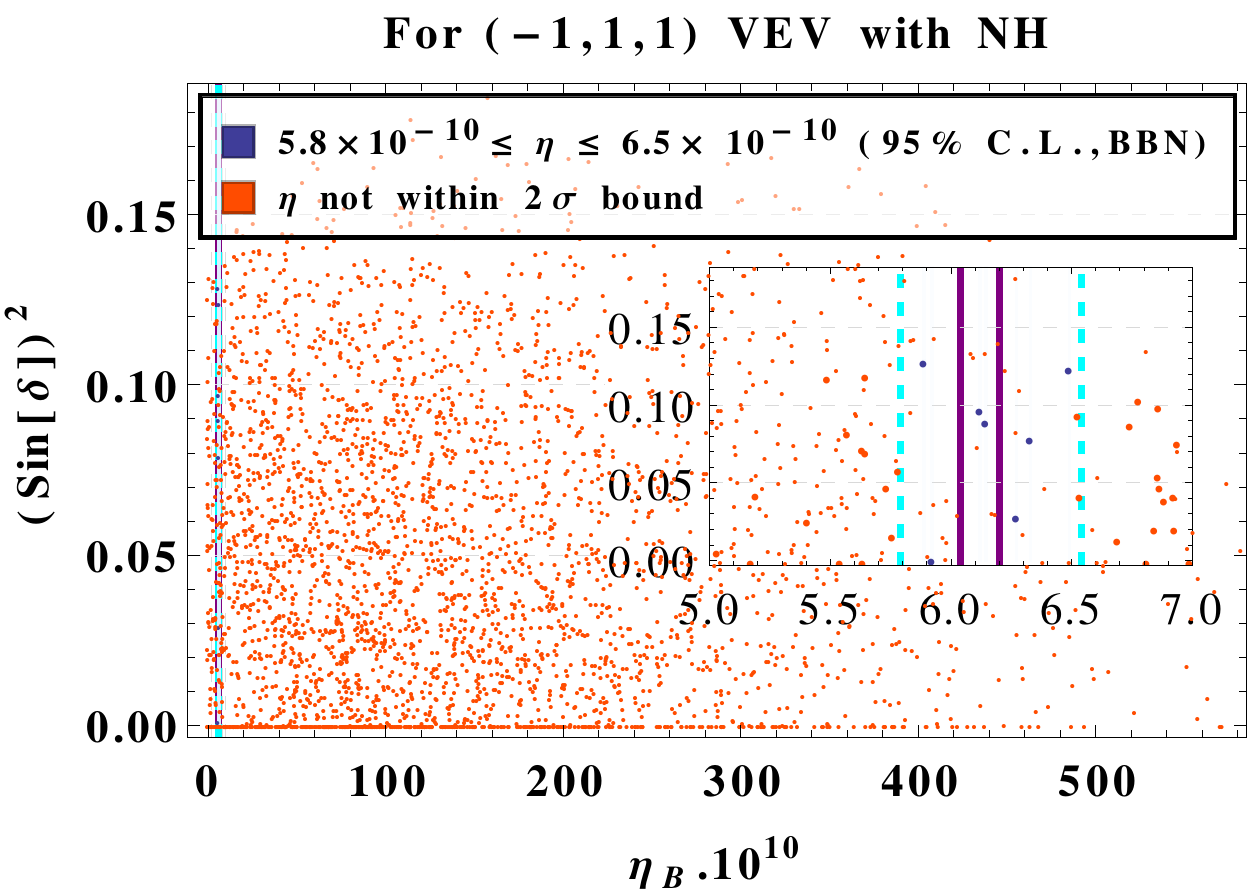}\label{fig4d}}
\caption{The plots (\ref{fig4a}), (\ref{fig4b}), (\ref{fig4c})  and (\ref{fig4d})  show the correlation between (i) $ \eta_{B}.10^{10} $ vs Log[$ \kappa $], (ii) $ \eta_{B}.10^{10} $ vs Log[$ \epsilon $], (iii) $ \eta_{B}.10^{10} $ vs $ m_{1} $ and (iv) $ \eta_{B}.10^{10} $ vs $(Sin \delta )^{2}$  for (-1,1,1) NH with $\dfrac{\upsilon^{\dagger}_{\rho}}{\Lambda}\sim 0.7$ respectively.}
\label{Fig-111 D}
\end{figure}
\begin{figure}[h]
\centering
\subfloat[]{\includegraphics[width=0.48\textwidth]{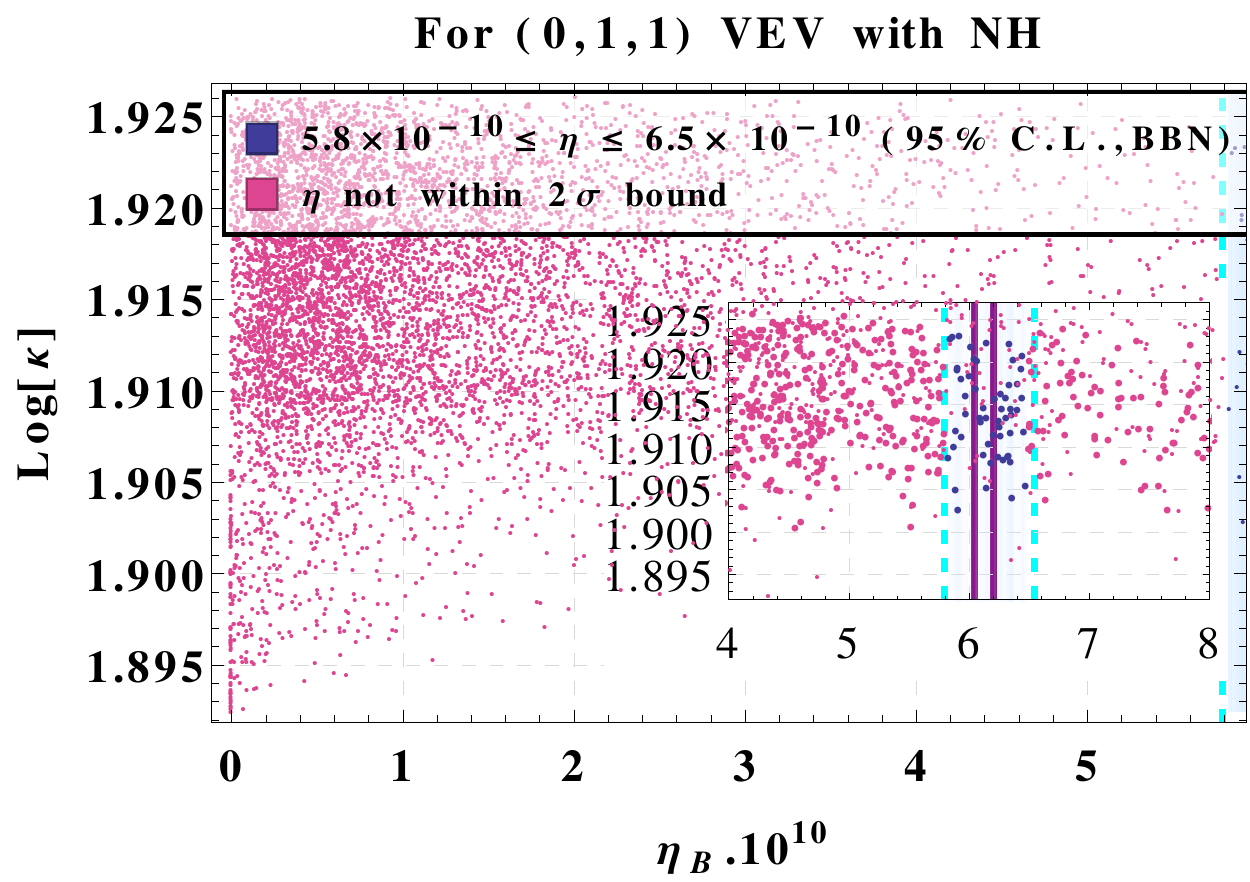}\label{fig6a}}
\qquad
\subfloat[]{\includegraphics[width=0.47\textwidth]{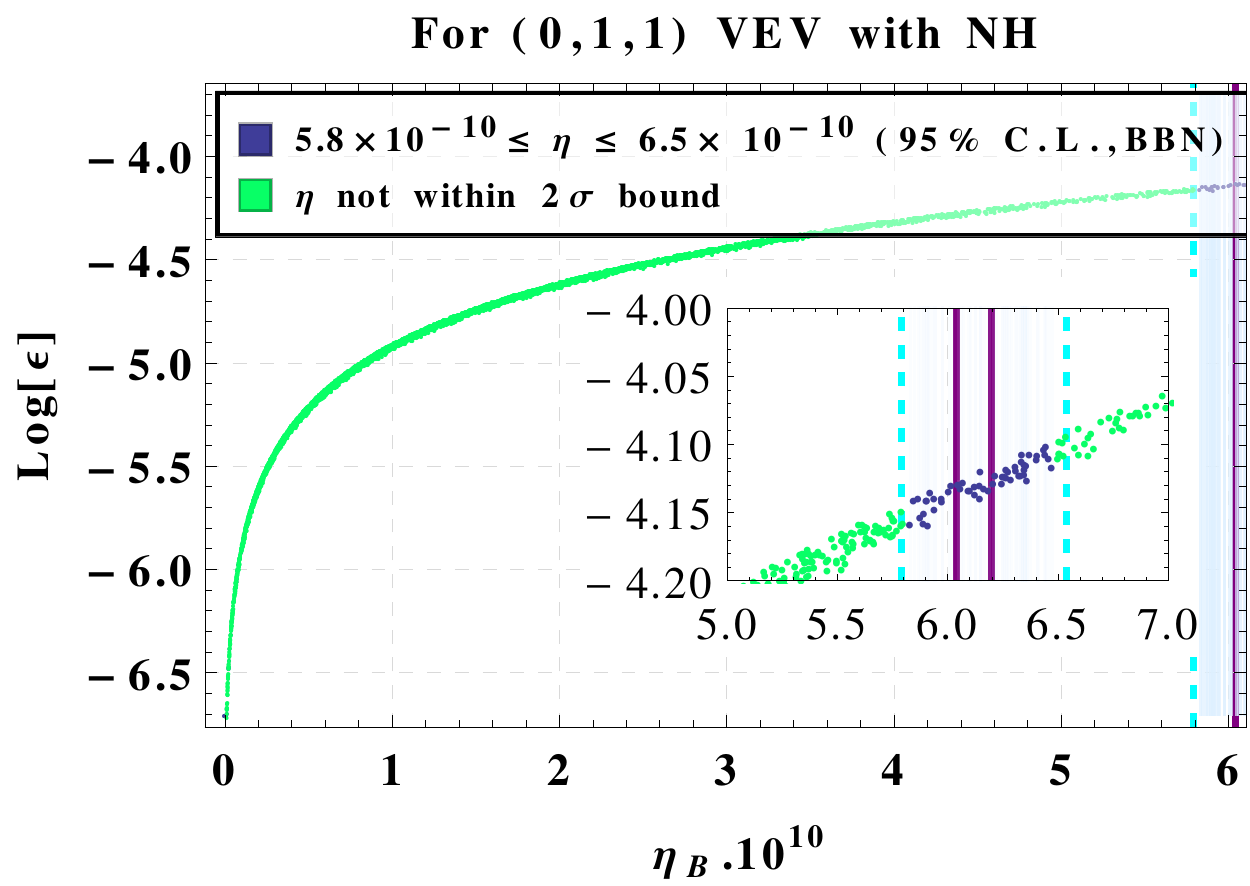}\label{fig6b}}
\qquad
\subfloat[]{\includegraphics[width=0.48\textwidth]{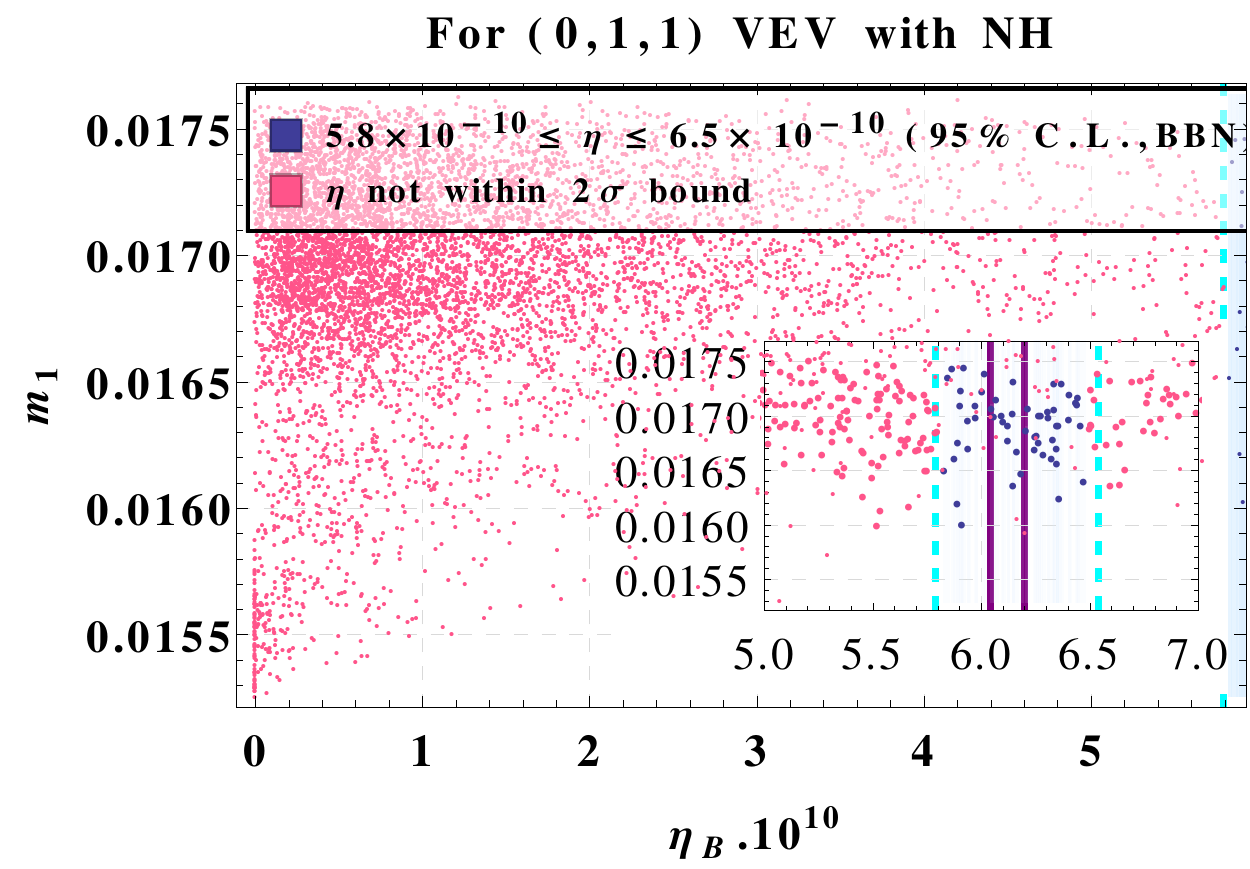}\label{fig6c}}
\qquad
\subfloat[]{\includegraphics[width=0.47\textwidth]{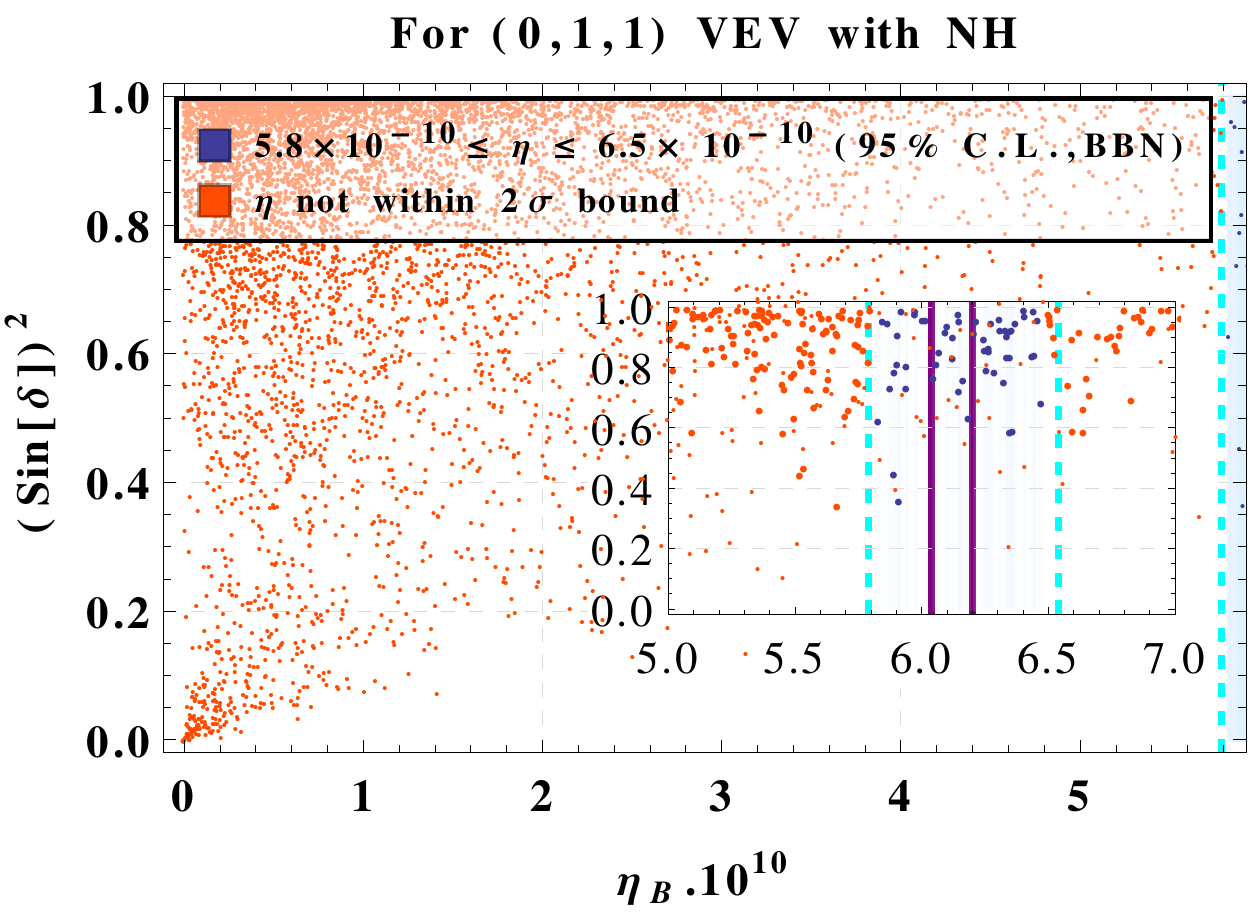}\label{fig6d}}
\caption{The plots (\ref{fig6a}), (\ref{fig6b}), (\ref{fig6c})  and (\ref{fig6d})  show the correlation between (i) $ \eta_{B}.10^{10} $ vs Log[$ \kappa $], (ii) $ \eta_{B}.10^{10} $ vs Log[$ \epsilon $], (iii) $ \eta_{B}.10^{10} $ vs $ m_{1} $ and (iv) $ \eta_{B}.10^{10} $ vs $(Sin \delta )^{2}$  for (0,1,1) NH with $\dfrac{\upsilon^{\dagger}_{\rho}}{\Lambda}\sim 0.5$ respectively.}
\label{Fig011 C}
\end{figure}
\begin{figure}[h]
\centering
\subfloat[]{\includegraphics[width=0.48\textwidth]{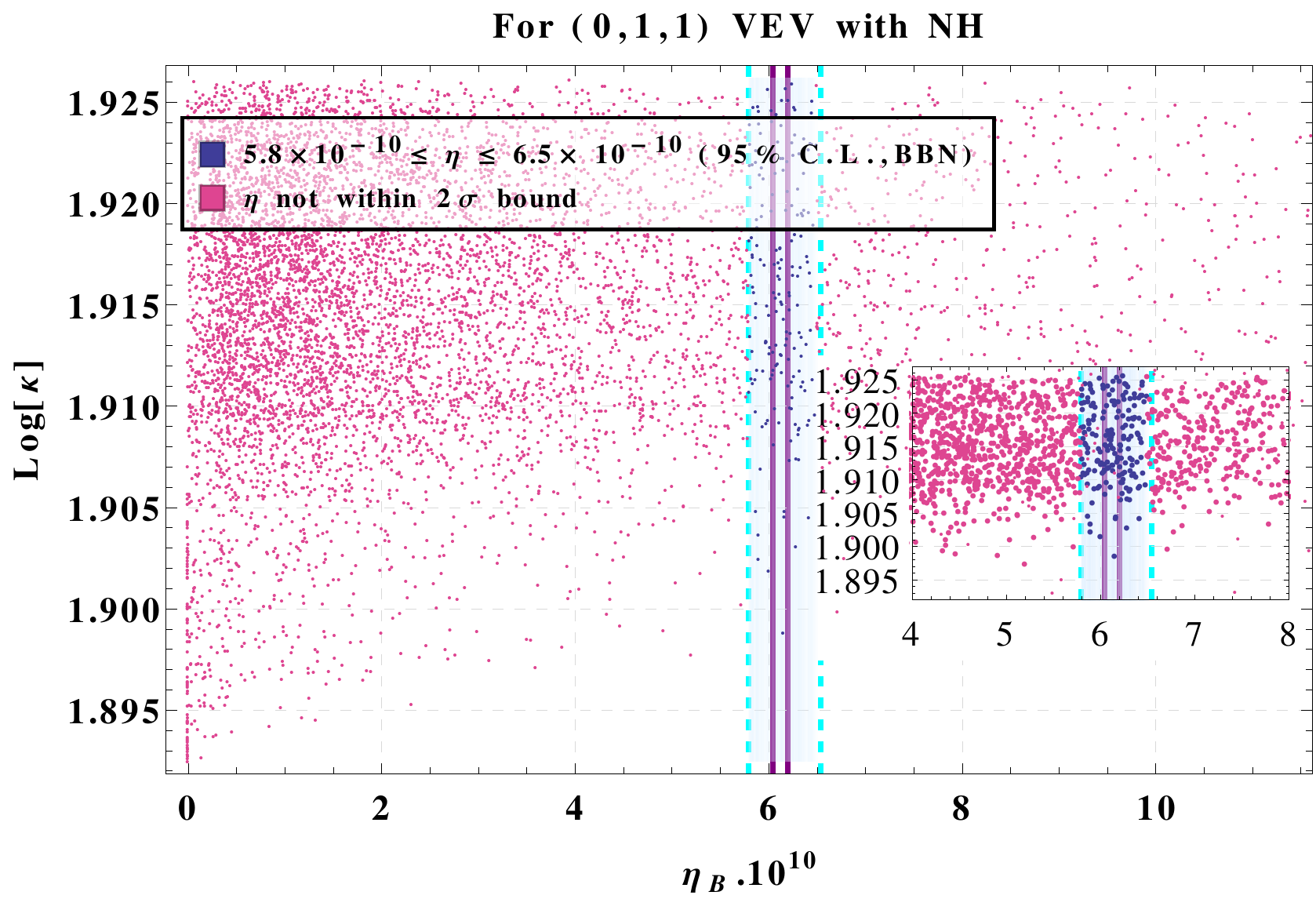}\label{fig7a}}
\qquad
\subfloat[]{\includegraphics[width=0.47\textwidth]{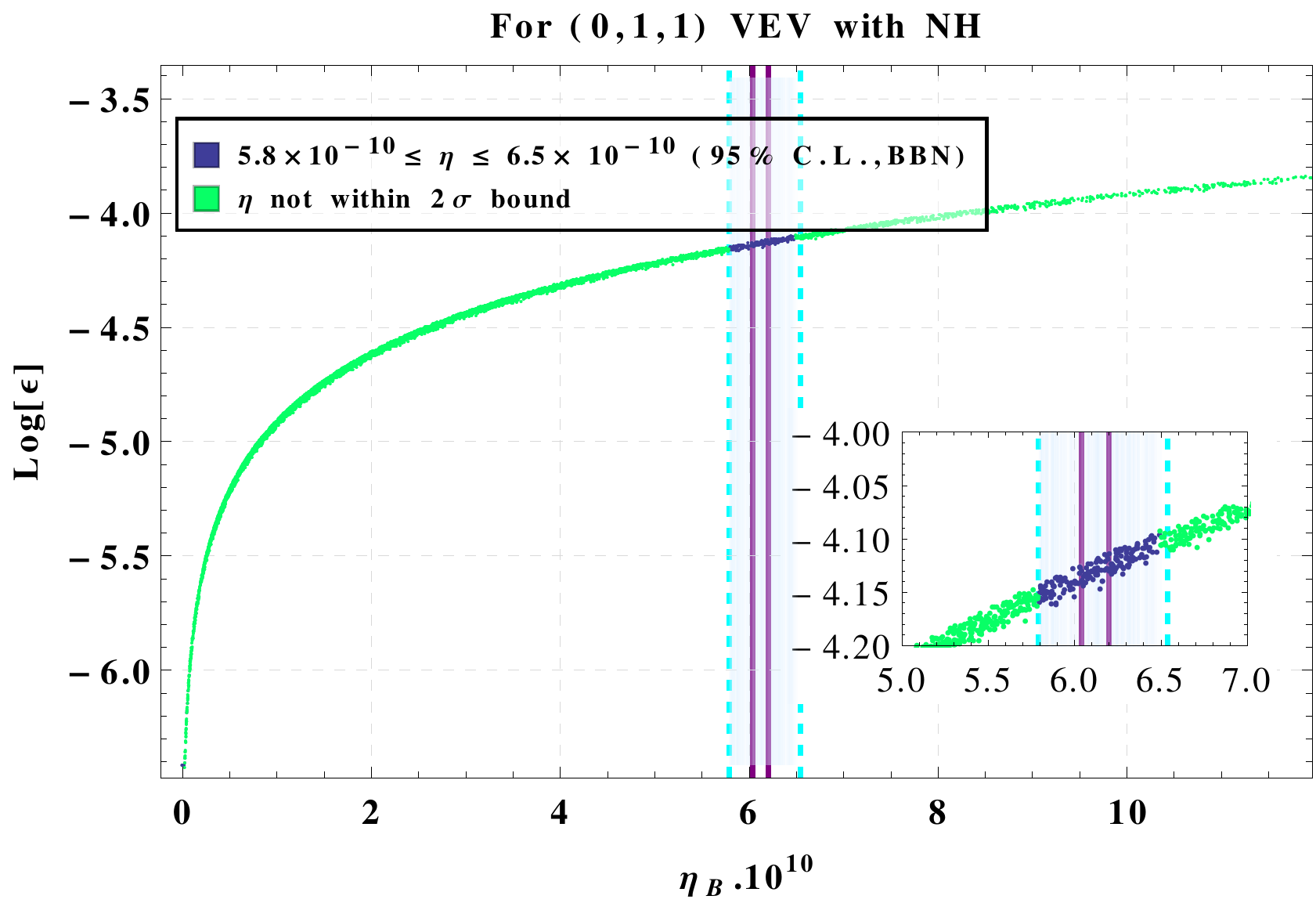}\label{fig7b}}
\qquad
\subfloat[]{\includegraphics[width=0.48\textwidth]{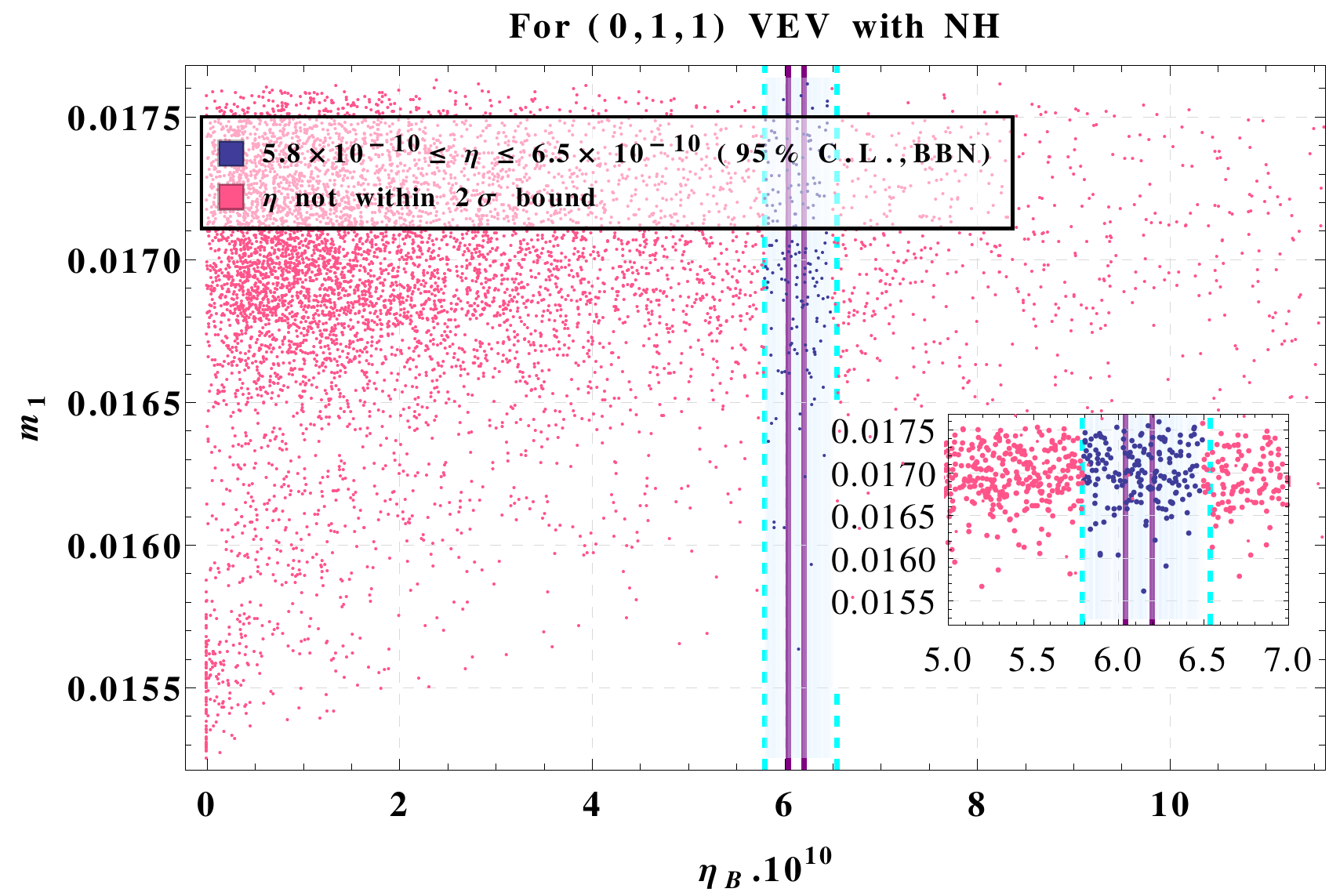}\label{fig7c}}
\qquad
\subfloat[]{\includegraphics[width=0.47\textwidth]{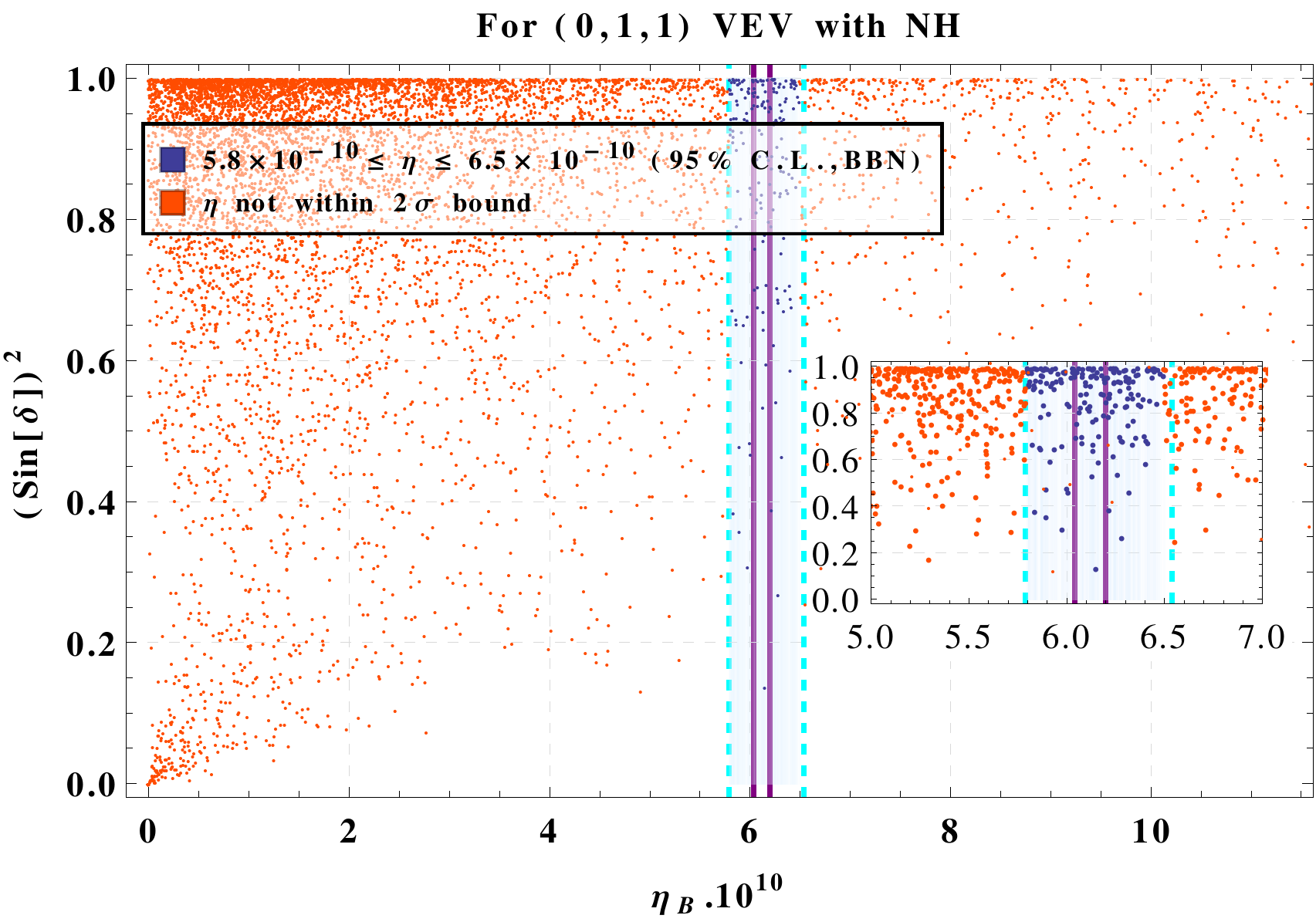}\label{fig7d}}
\caption{The plots (\ref{fig7a}), (\ref{fig7b}), (\ref{fig7c})  and (\ref{fig7d})  show the correlation between (i) $ \eta_{B}.10^{10} $ vs Log[$ \kappa $], (ii) $ \eta_{B}.10^{10} $ vs Log[$ \epsilon $], (iii) $ \eta_{B}.10^{10} $ vs $ m_{1} $ and (iv) $ \eta_{B}.10^{10} $ vs $(Sin \delta )^{2}$  for (0,1,1) NH with $\dfrac{\upsilon^{\dagger}_{\rho}}{\Lambda}\sim 0.7$ respectively.}
\label{Fig011 D}
\end{figure}
\begin{figure}[h]
\centering
\subfloat[]{\includegraphics[width=0.48\textwidth]{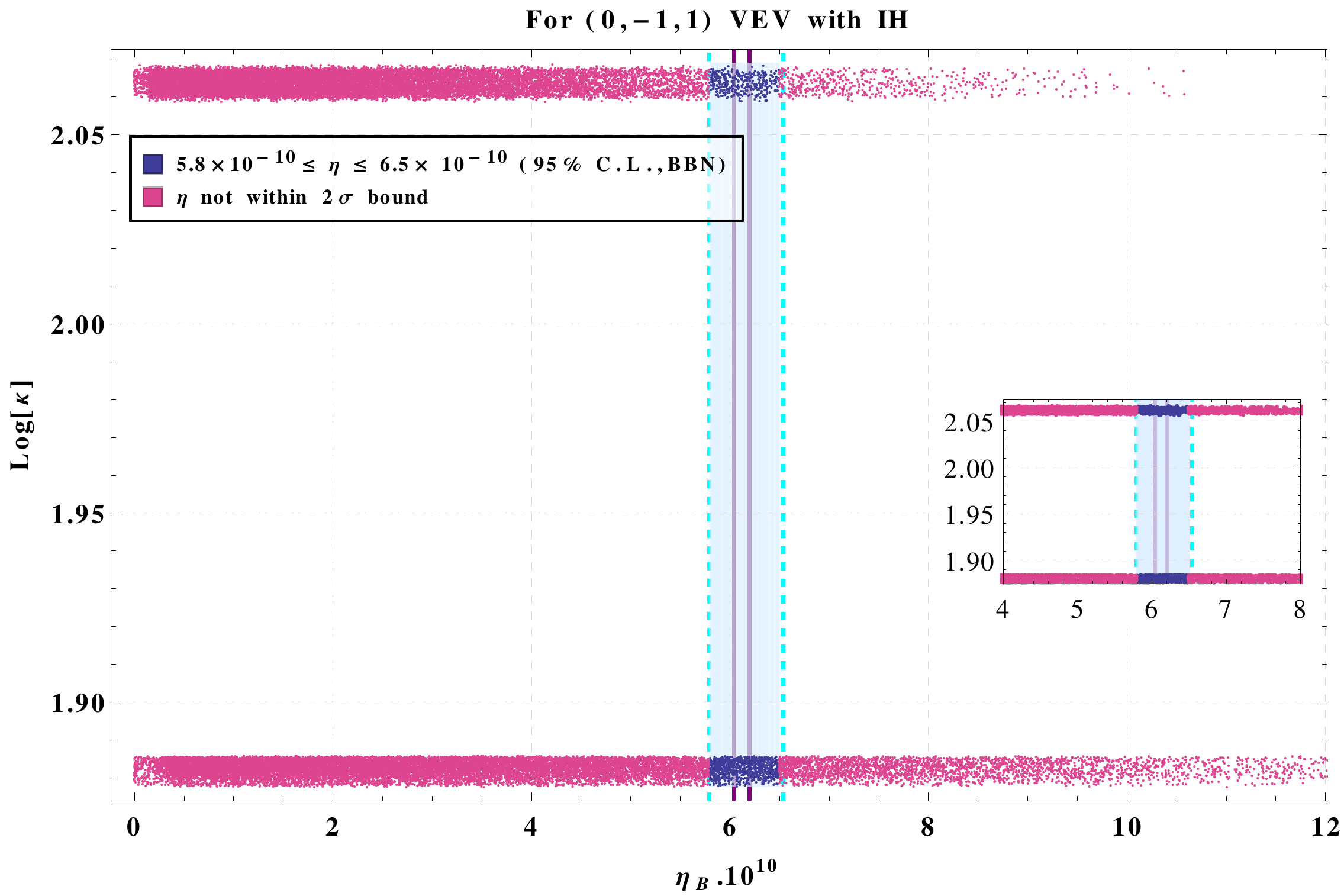}\label{fig8a}}
\qquad
\subfloat[]{\includegraphics[width=0.47\textwidth]{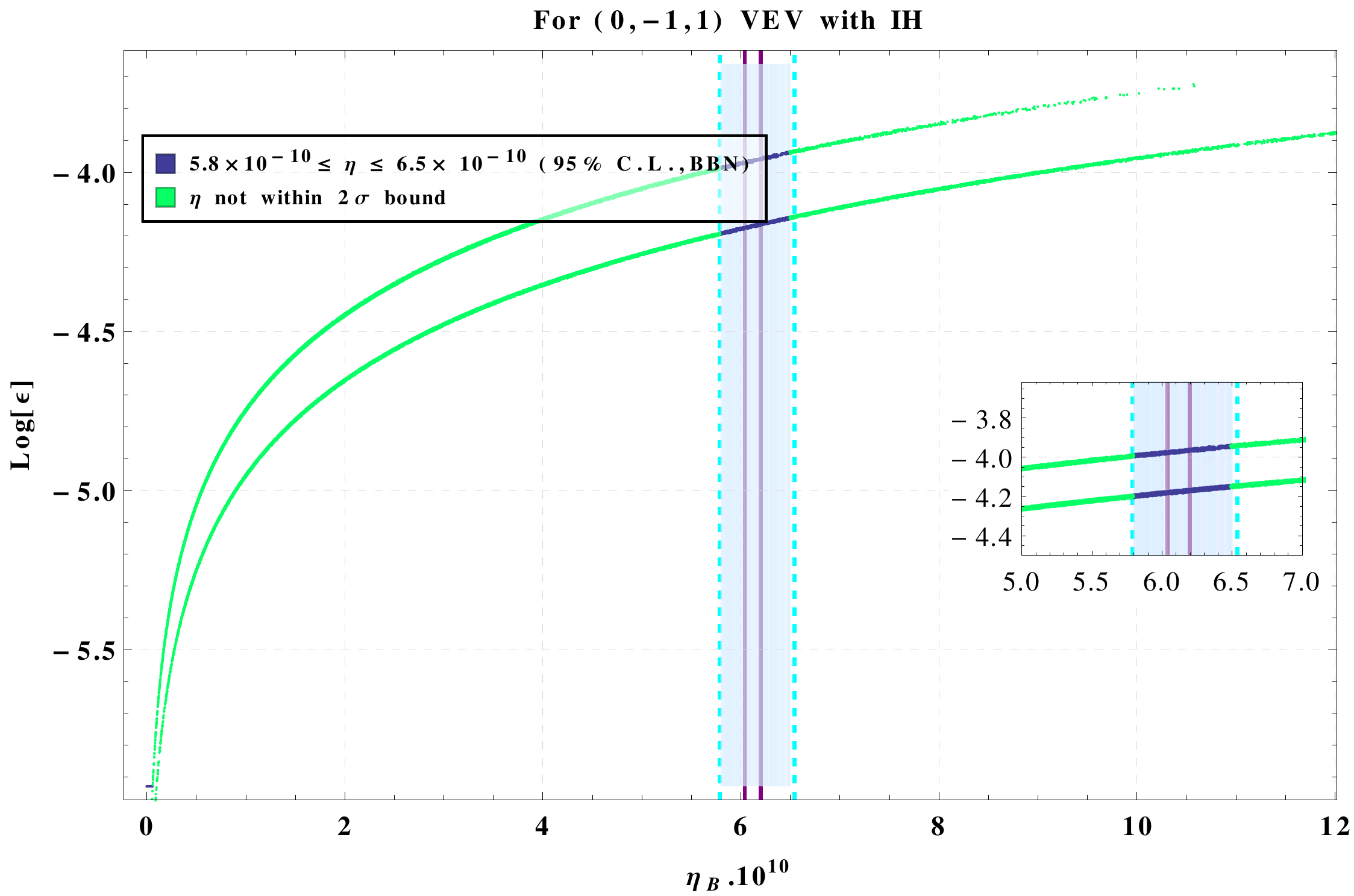}\label{fig8b}}
\qquad
\subfloat[]{\includegraphics[width=0.48\textwidth]{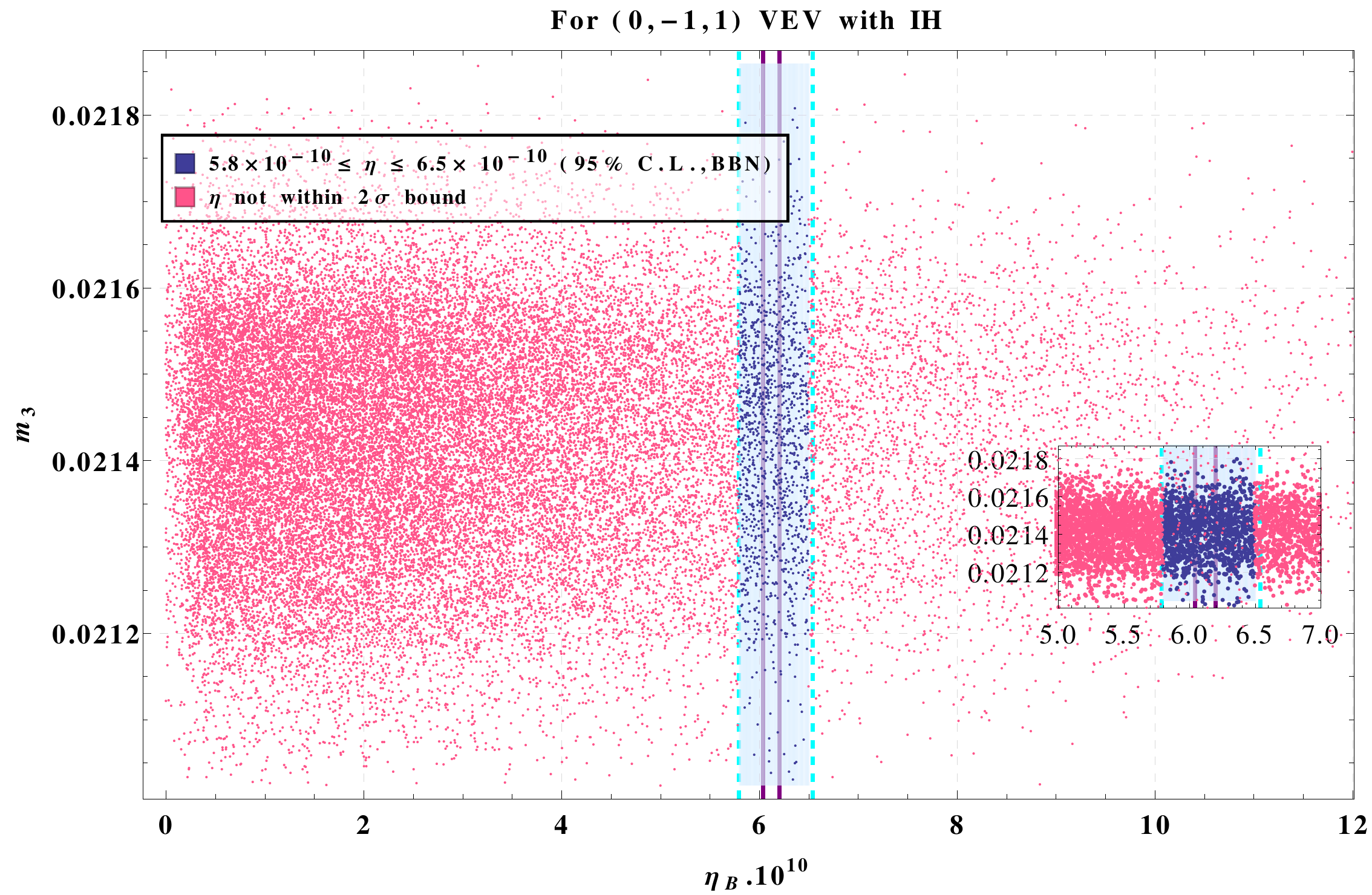}\label{fig8c}}
\qquad
\subfloat[]{\includegraphics[width=0.47\textwidth]{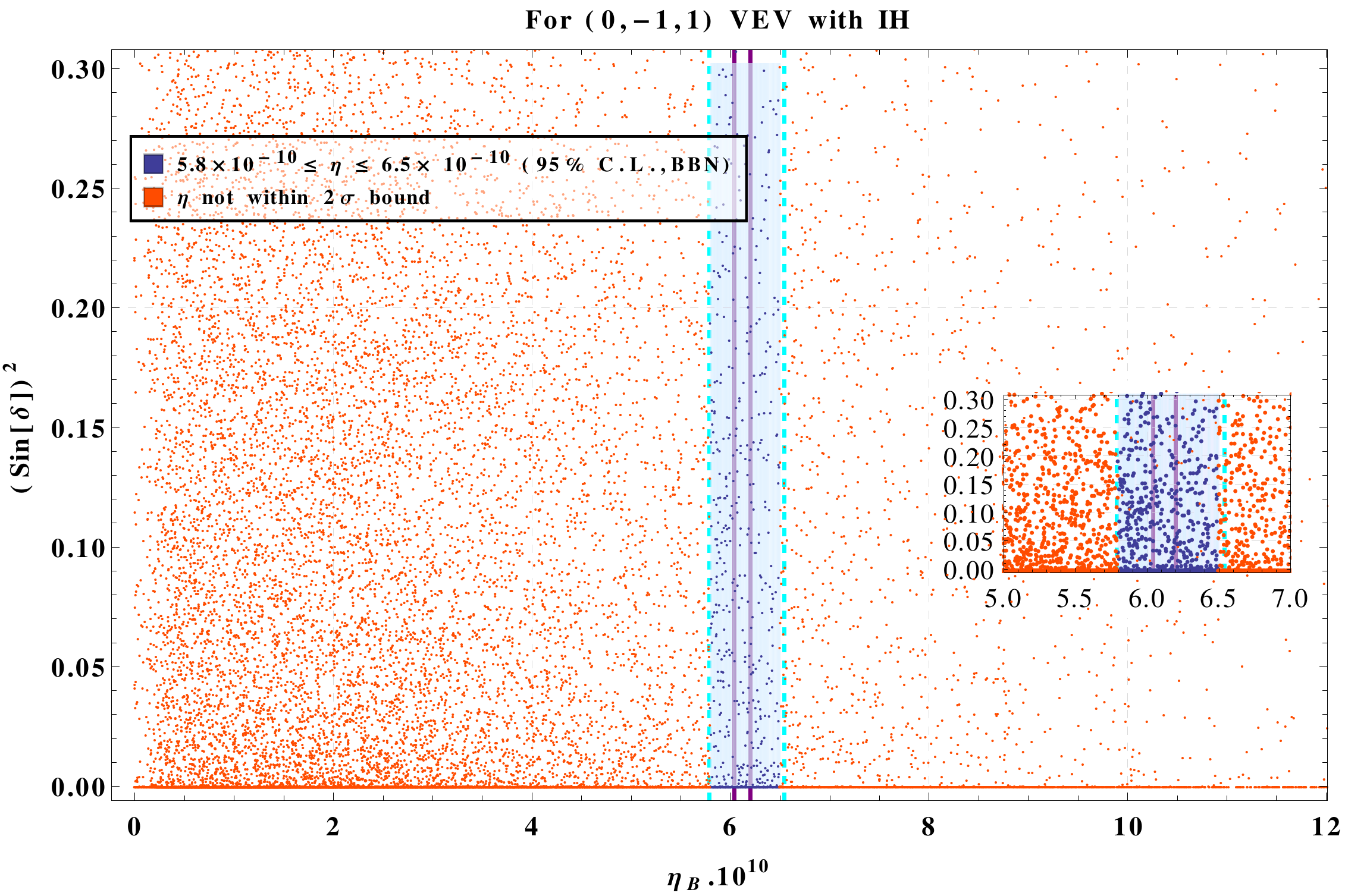}\label{fig8d}}
\caption{The plots (\ref{fig8a}), (\ref{fig8b}), (\ref{fig8c})  and (\ref{fig8d})   show the correlation between (i) $ \eta_{B}.10^{10} $ vs Log[$ \kappa $], (ii) $ \eta_{B}.10^{10} $ vs Log[$ \epsilon $], (iii) $ \eta_{B}.10^{10} $ vs $ m_{3} $ and (iv) $ \eta_{B}.10^{10} $ vs $(Sin \delta )^{2}$  for (0,1,-1) IH with $\dfrac{\upsilon^{\dagger}_{\rho}}{\Lambda}\sim 0.3 $ respectively.}
\label{01-1 B}
\end{figure}
\begin{figure}[h]
\centering
\subfloat[]{\includegraphics[width=0.48\textwidth]{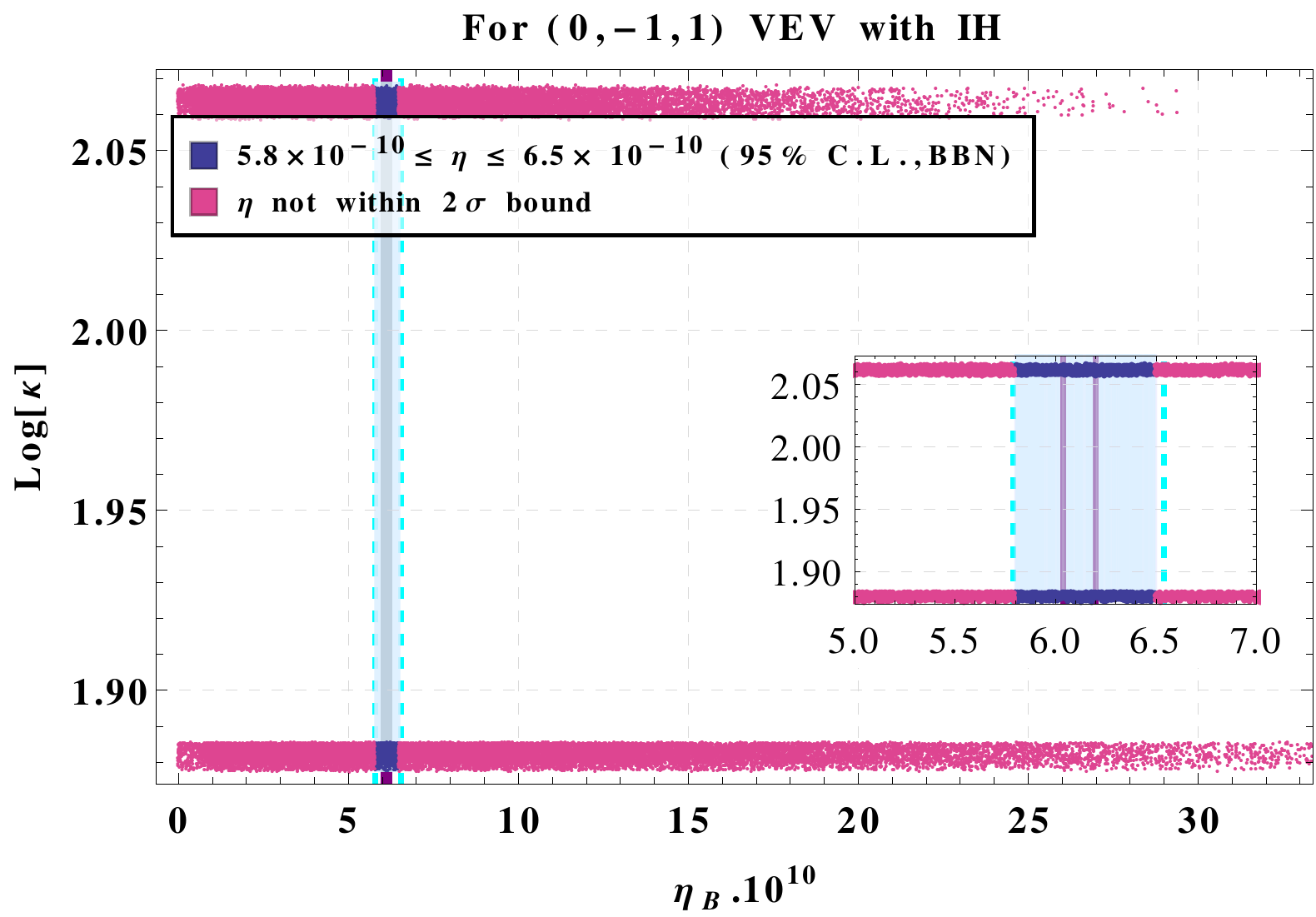}\label{fig9a}}
\qquad
\subfloat[]{\includegraphics[width=0.47\textwidth]{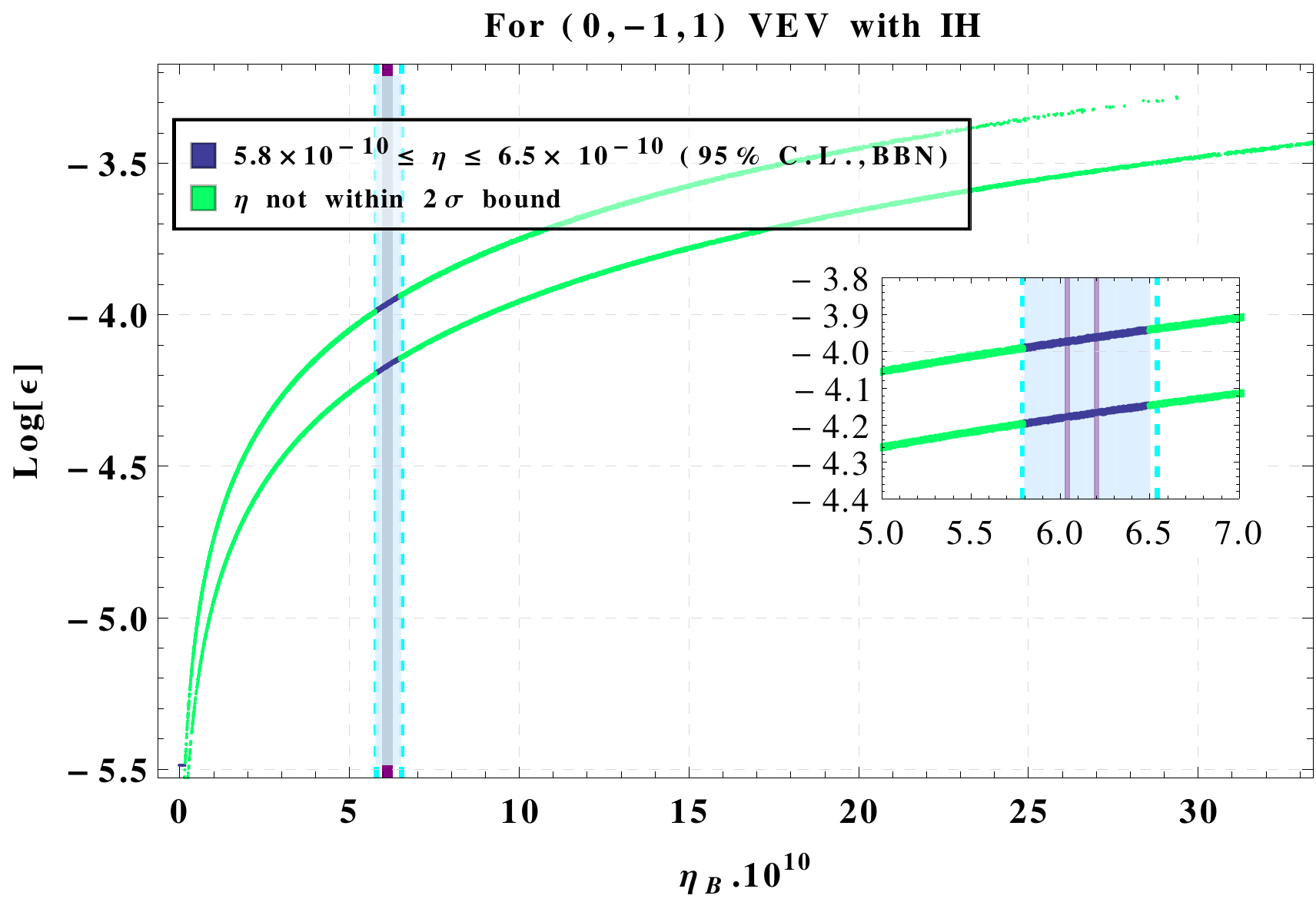}\label{fig9b}}
\qquad
\subfloat[]{\includegraphics[width=0.48\textwidth]{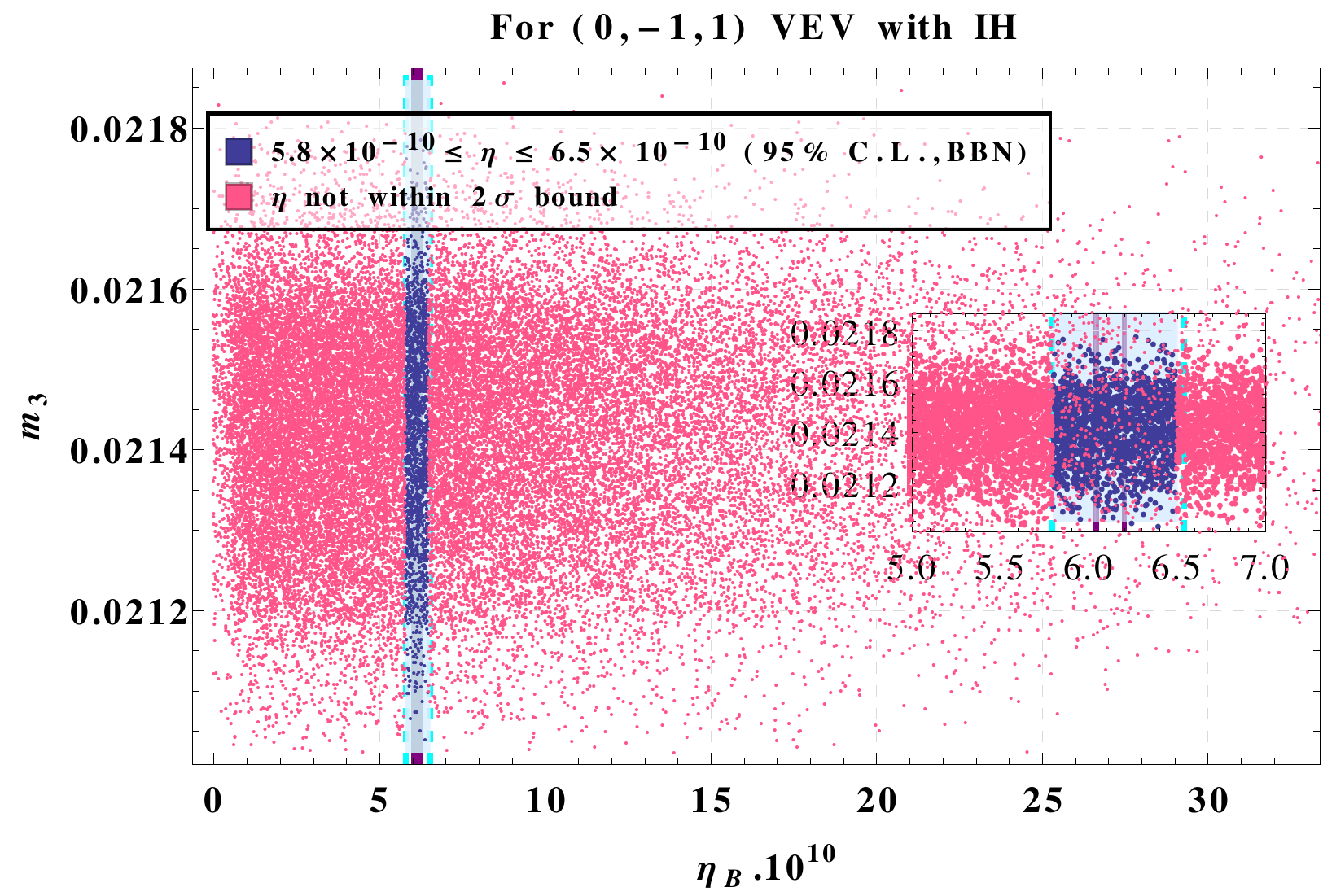}\label{fig9c}}
\qquad
\subfloat[]{\includegraphics[width=0.47\textwidth]{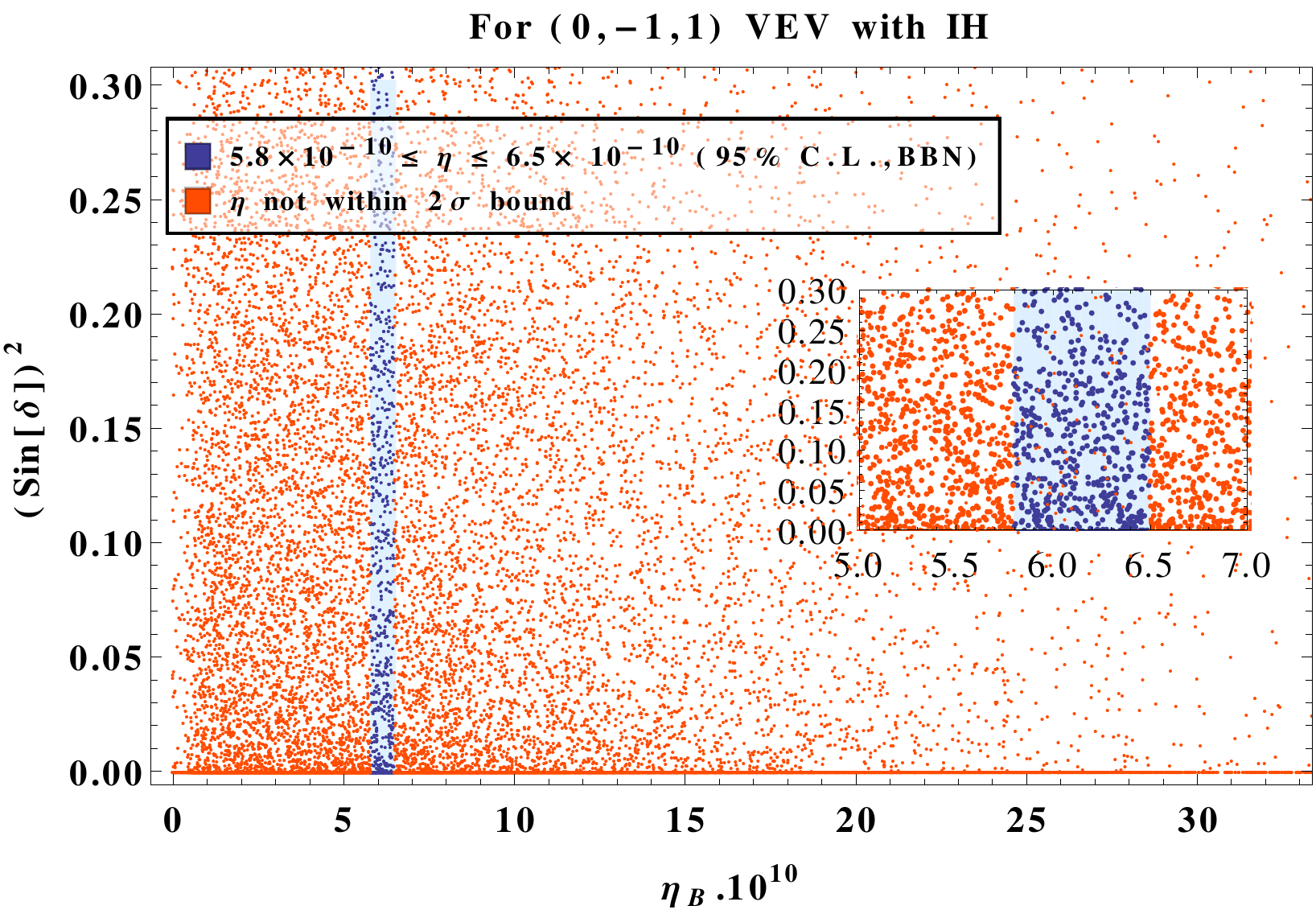}\label{fig9d}}
\caption{The plots (\ref{fig9a}), (\ref{fig9b}), (\ref{fig9c})  and (\ref{fig9d})   show the correlation between (i) $ \eta_{B}.10^{10} $ vs Log[$ \kappa $], (ii) $ \eta_{B}.10^{10} $ vs Log[$ \epsilon $], (iii) $ \eta_{B}.10^{10} $ vs $ m_{3} $ and (iv) $ \eta_{B}.10^{10} $ vs $(Sin \delta )^{2}$  for (0,1,-1) IH with $\dfrac{\upsilon^{\dagger}_{\rho}}{\Lambda}\sim 0.5 $ respectively.}
\label{01-1 C}
\end{figure}
\begin{figure}[h]
\centering
\subfloat[]{\includegraphics[width=0.48\textwidth]{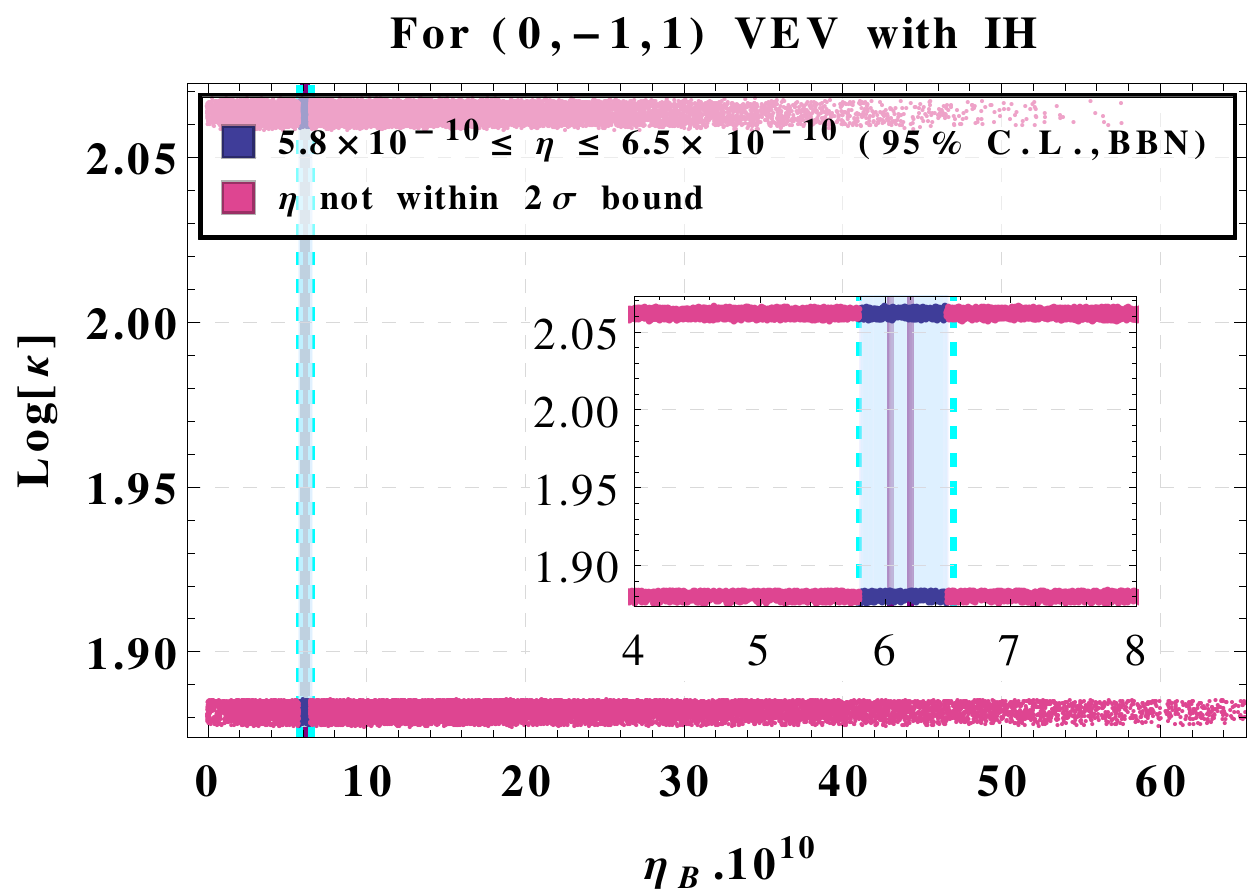}\label{fig10a}}
\qquad
\subfloat[]{\includegraphics[width=0.47\textwidth]{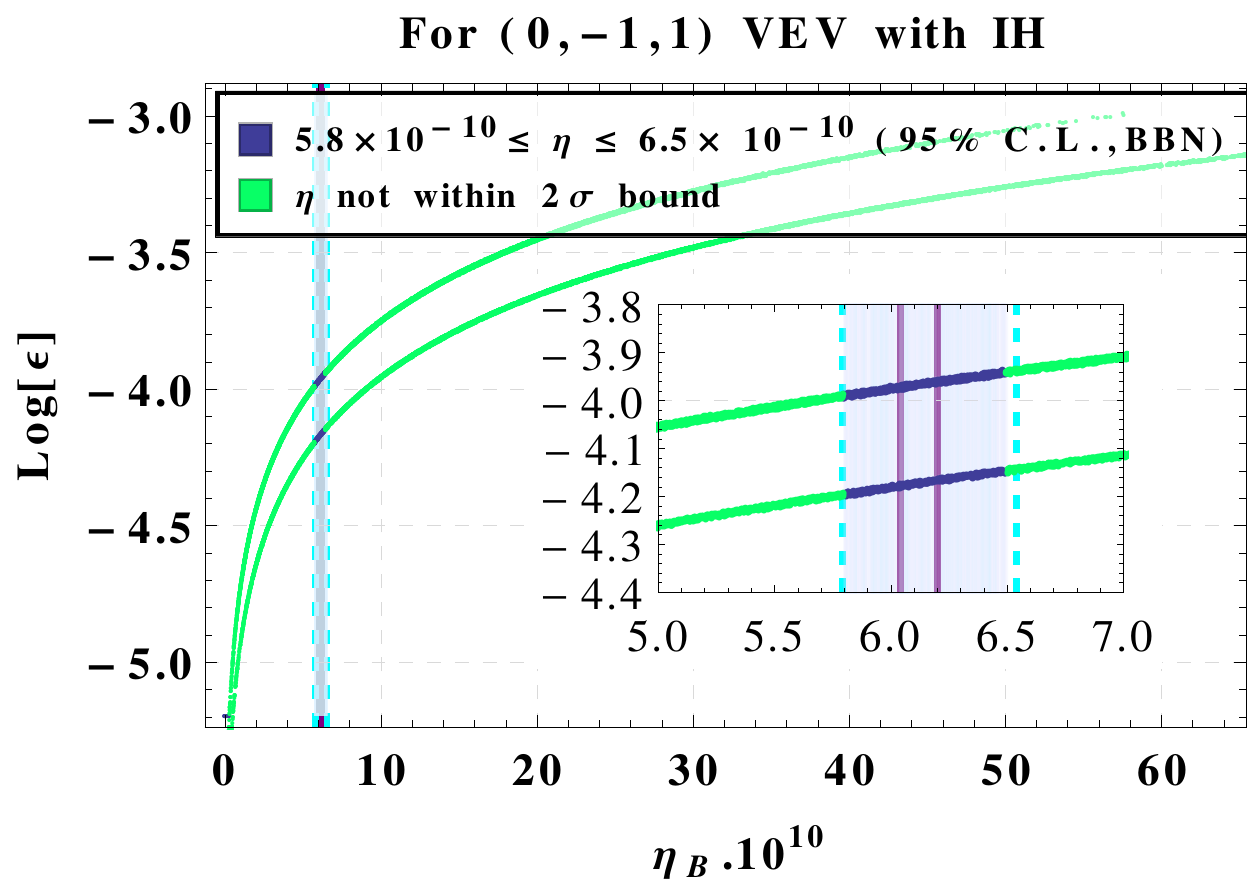}\label{fig10b}}
\qquad
\subfloat[]{\includegraphics[width=0.48\textwidth]{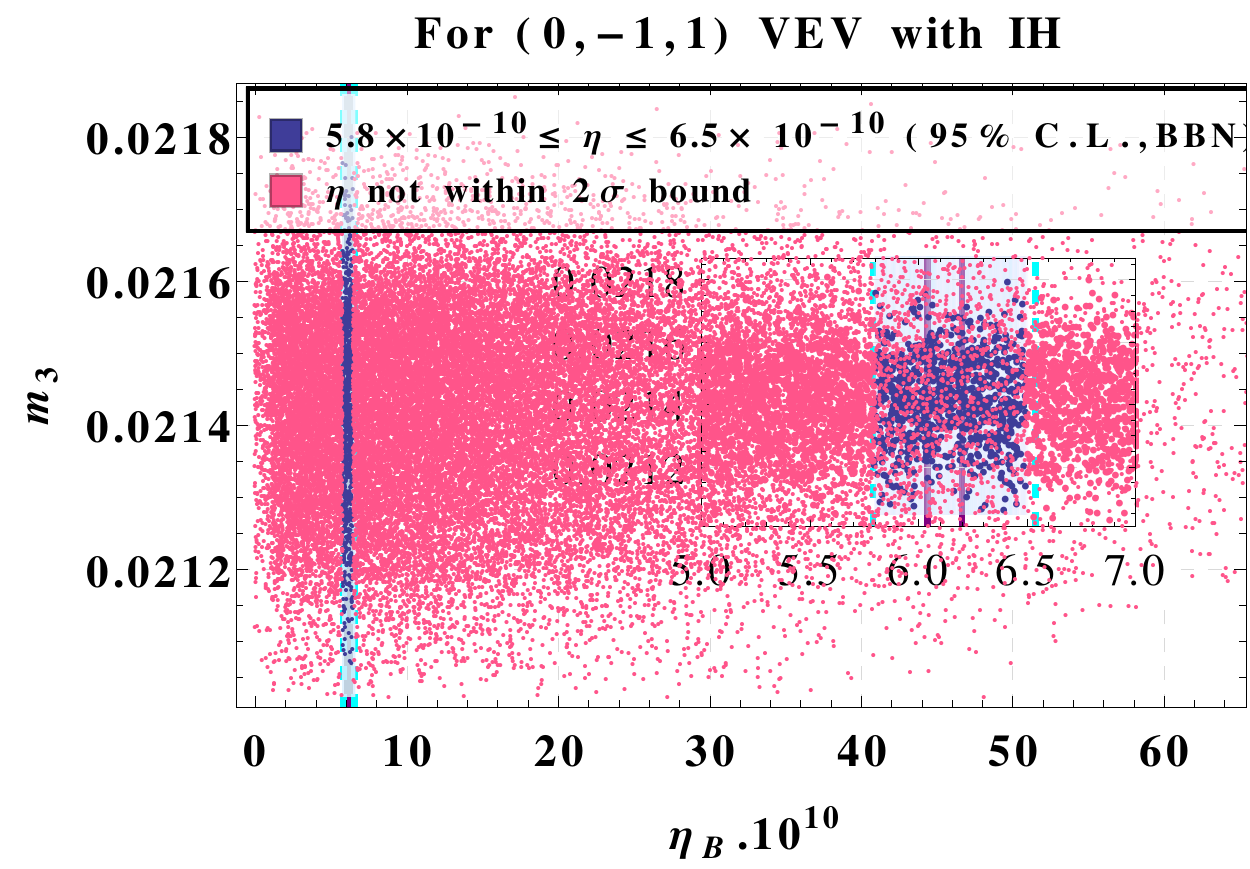}\label{fig10c}}
\qquad
\subfloat[]{\includegraphics[width=0.47\textwidth]{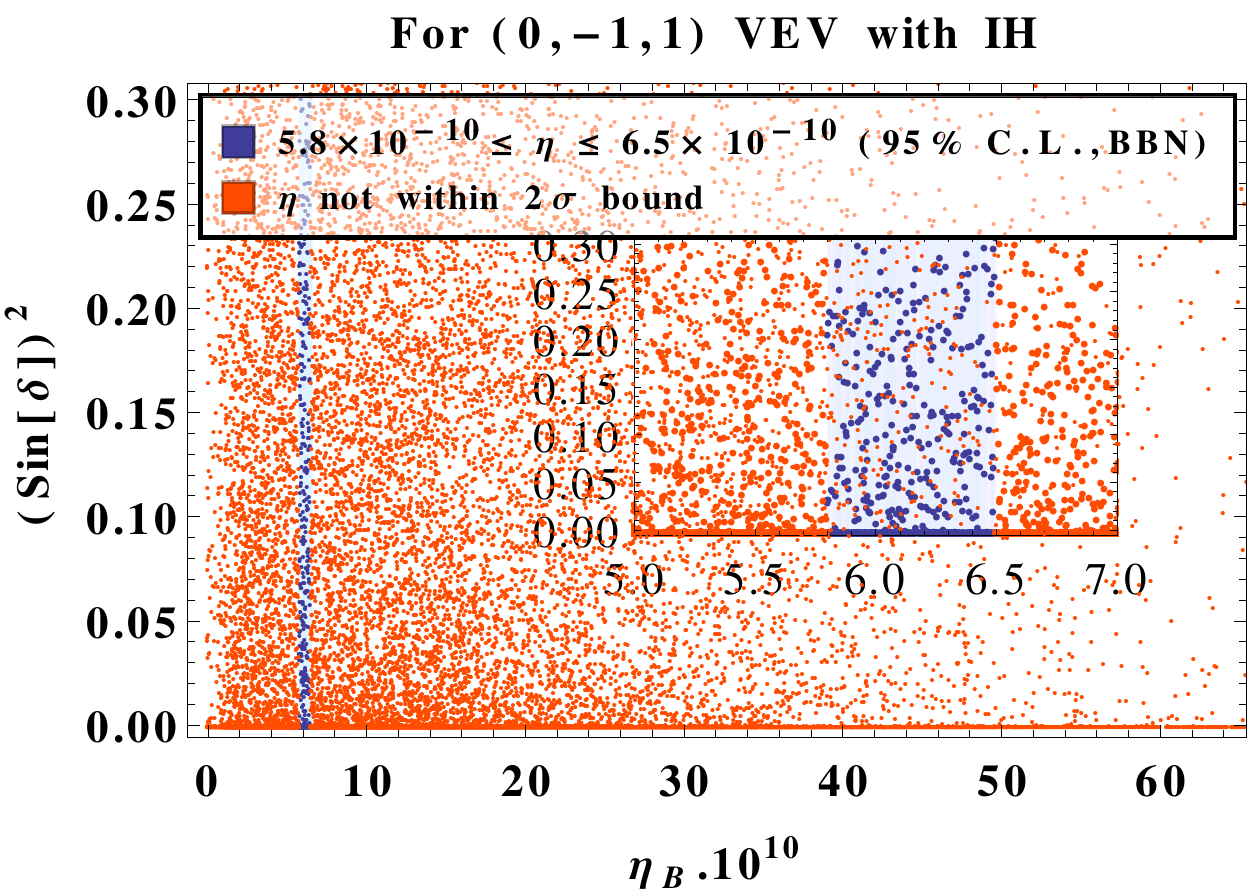}\label{fig10d}}
\caption{The plots (\ref{fig10a}), (\ref{fig10b}), (\ref{fig10c})  and (\ref{fig10d})   show the correlation between (i) $ \eta_{B}.10^{10} $ vs Log[$ \kappa $], (ii) $ \eta_{B}.10^{10} $ vs Log[$ \epsilon $], (iii) $ \eta_{B}.10^{10} $ vs $ m_{3} $ and (iv) $ \eta_{B}.10^{10} $ vs $(Sin \delta )^{2}$  for (0,1,-1) IH with $\dfrac{\upsilon^{\dagger}_{\rho}}{\Lambda}\sim 0.7 $ respectively.}
\label{01-1 D}
\end{figure}
It is seen from Table (\ref{Tab:lept}) that in all the cases listed, we get baryon asymmetry of the universe within the allowed region of $ \eta_{B} $ \cite{Fields:2014uja}, except for (0,1,1) in NH and (0,1,-1) in IH for $ \dfrac{\upsilon^{\dagger}_{\rho}}{\Lambda} \sim 0.1 $.  The decay parameter $ \kappa $ is found to be greater than 1 for all three allowed cases. This indicates a strong washout scenario where the heavy Majorana neutrinos even at temperature $ T < M_{1}$ do not decouple from the thermal bath immediately, and the forward and inverse decay reactions thermalise very quickly at $ T \sim M_{1} $ \cite{Biondini:2016hhn}. This removes any pre-existing asymmetry before leptogenesis occurs. Thus, our model indicate towards strong washout  leading to highly predictive resonant leptogenesis that doesn't depend on the initial conditions \cite{Biondini:2016hhn}. Although the generation of matter-antimatter asymmetry at $ T< M_{1} $ through the decay of the lightest Majorana neutrinos is applicable for both the weak and strong washout scenarios, however, it is worth mentioning that in the case of strong washout, the lepton asymmetry begins immediately after $ T \sim M_{1} $.  Whereas, for weak washout, the asymmetry is gradually generated at later times \cite{Biondini:2016hhn}. From results presented in Table (\ref{Tab:lept}) and Figs. (\ref{Fig-111 A}- \ref{01-1 D}), we conclude that for $ 0.1 \leq  \dfrac{\upsilon^{\dagger}_{\rho}}{\Lambda} \leq 0.7 $, the BAU lies within the experimental bounds which imply that resonant leptogenesis can be successfully enhanced in our inverse seesaw scenario. Leptogenesis can have a sizeable parameter space, as seen from our results in Figs (\ref{Fig-111 A}- \ref{01-1 D}) with a total of six yet unknown parameters. These variables are the lightest neutrino mass $m_1$, the Dirac and Majorana phases of the PMNS matrix and the lepton asymmetry and baryon number density. \\
\\
By carefully examining the plots (\ref{Fig-111 A}- \ref{01-1 D}), it is evident that \\
\\
1. With increasing value of $\dfrac{\upsilon^{\dagger}_{\rho}}{\Lambda}$, the number of points in the allowed region of $\eta_{B}$ increases  for (0,1,1)/NH and (0,-1,1)/IH, and decreases  for (-1,1,1)/NH .\\
\\
2.  There are no points in the allowed region of $\eta_B$,  for $\dfrac{\upsilon^{\dagger}_{\rho}}{\Lambda} =$0.1 and 0.3,  for (0,1,1)/NH and (-1,1,1)/IH, as well as $\dfrac{\upsilon^{\dagger}_{\rho}}{\Lambda} =0.1$ for (-1,1,1)/NH. \\
\\
3. Hence, we can say that the value of BAU depends on the decay parameter $ \kappa $,  triplet flavon alignment and the CP asymmetry. \\
\\ 
4. From the plot of light neutrino mass versus baryon number density,  in figures (\ref{fig1c},  \ref{fig2c}, \ref{fig3c}, \ref{fig4c}, \ref{fig6c}, \ref{fig7c}, \ref{fig8c}, \ref{fig9c}, \ref{fig10c}), it is observed that range of this mass has a dependence on triplet flavon VEV alignment.
\\
\\
Hence, in future, if the mass hierarchy and the absolute value of mass of light neutrinos is measured with precision,  the results presented in this work can throw some light on scale of flavor symmetry breaking as well as favored VEV alignment of the triplet flavon of $A_4$ symmetry. Thus, BAU can be connected with dynamics of flavor symmetry.

\section{Summary and Conclusion}
The baryon asymmetry of the universe may be explained by the mechanism of leptogenesis, which states that the excess of matter over antimatter in the universe is due to the decay of the lightest Majorana neutrinos. In order for this mechanism to be able to produce the observed baryon asymmetry of the universe, the neutrino oscillation parameters must lie within a  certain range. In this study, we investigated the effect of the neutrino oscillation parameters on the baryon asymmetry of the universe in the context of resonant leptogenesis,  and the BAU lies within its experimental bounds. We considered our previously developed ISS model with $A_4$ flavor symmetry. We found that the BAU of the universe is sensitive to the lightest neutrino  mass, which must lie within a certain range in order to achieve the observed asymmetry. It is  also observed that the BAU is significantly affected by the mass splitting between the two heaviest neutrinos and  CP-violating phases.  \\
\\
We observed that the decay parameter $ \kappa $ is found to be greater than 1 for all three allowed cases.  It indicates that the theory of resonant leptogenesis follows a strong washout scenario where the heavy Majorana neutrinos even at temperature $ T < M_{1}$ don't decouple from the thermal bath immediately and the forward and inverse decay reactions thermalise very quickly. It removes any pre-existing asymmetry before leptogenesis occurs, leading to highly predictive resonant leptogenesis that doesn't depend on the initial conditions.  BAU in our model is found to lie within currently allowed limits for a large parameter space with six unknown parameters, including the lightest neutrino mass $m_1$, Dirac and Majorana phases of the PMNS matrix, lepton asymmetry and baryon number density. The permitted mixing angles of the heavy neutrinos are constrained by the requirement of recreating both the baryon asymmetry of the universe and the light neutrino masses through our inverse seesaw model.\\
\\
From these results, we find that more favorable cases from BAU point of view (VEV alignment  of triplet flavon and scale of symmetry breaking of $A_4$ flavor symmetry, and other parameters) are as follows:\\
\begin{table}[h]
        \centering
        \begin{tabular}{| l | c | c | c |}
\hline 
VEV of $ \Phi_{s} $  & Scale $ \dfrac{\upsilon^{\dagger}_{\rho}}{\Lambda} $ & Favoured ($ \checkmark $) / Disfavoured ($ \times $) \\ 
\hline
  & 0.1 & $ \checkmark $ \\ 
  \cline{2-3}
  & 0.3  & $ \checkmark $ \\
   \cline{2-3}
(-1,1,1) in NH   & 0.5  &$ \checkmark $ \\
   \cline{2-3}
   & 0.7  & $ \checkmark $ \\
\hline 
\hline
 & 0.1  & $ \times $ \\
  \cline{2-3}
& 0.3 &  $ \times $ \\ 
  \cline{2-3}
(0,1,1) in NH   & 0.5  & $ \checkmark $\\
   \cline{2-3}
    & 0.7  & $ \checkmark $\\
\hline 
\hline
 & 0.1  & $ \times $ \\
  \cline{2-3}
& 0.3 &  $ \checkmark $\\ 
  \cline{2-3}
(0,-1,1) in IH   & 0.5  & $ \checkmark $\\
   \cline{2-3}
    & 0.7  & $ \checkmark $\\
\hline 
\end{tabular}
      \caption{Favoured and disfavoured flavor symmetry breaking scale for allowed VEV alignment of triplet scalar flavon for which BAU computed in our model lies within its current 3$ \sigma $ range.}
        \label{tab:final}
             \end{table}\\     
Upon analyzing Table (\ref{tab:final}) and Figures (1-10), it is evident that a decrease in the value of $ \dfrac{\upsilon^{\dagger}_{\rho}}{\Lambda} $ results in a lower probability of locating a correlation between the model parameters and BAU parameters. The VEV (-1,1,1) NH has a higher probability of probing the BAU for $ \dfrac{\upsilon^{\dagger}_{\rho}}{\Lambda} $ values ranging from $ (0.1 \rightarrow 0.7) $, while (0,1,1) NH has the lowest probability of probing the BAU. Also, an increase in the value of $ \dfrac{\upsilon^{\dagger}_{\rho}}{\Lambda} $ leads to an increase in the value of $ \eta_{B}.10^{10} $.
\\
\\
We would like to highlight that our analysis can give important insights in understanding the role of the model parameters to find BAU and its parameters. By identifying the VEVs that have a higher probability of probing the BAU, researchers can focus on the specific VEV alignment where the BAU can be highly predicted. Furthermore, the observed increase in $ \eta_{B}.10^{10} $ with an increase in $ \dfrac{\upsilon^{\dagger}_{\rho}}{\Lambda} $ highlights the importance of considering these parameters in future studies.          
In conclusion, our model has successfully predicted the BAU through resonant leptogenesis, and it is found that  the BAU depends on VEV alignment of the triplet flavon and scale of symmetry breaking of the $A_4$ flavor symmetry, the lightest neutrino mass, the mass splitting between the two heaviest neutrinos and the CP-violating phases. Thus, one can pinpoint their favored values, and can obtain some information on hitherto unknown dynamics of flavor symmetry. Hence, our model is testable when all the light neutrino oscillation parameters will be fixed in future, and it has been successfully able to explain several phenomenon like light neutrino oscillation, charged lepton flavor violation and BAU, in a unified manner. The findings reported in this work will provide a benchmark for future  studies on neutrino parameters and BAU in the framework of resonant leptogenesis.

\section*{Acknowledgment}
The authors are grateful to Professor M. P. Bora of the Department of Physics at Gauhati University for his assistance in creating a figure. 

\end{document}